\documentclass[twocolumn,10pt,aps,pra,showpacs,superscriptaddress]{revtex4-1}

\usepackage{amsmath,amssymb,mathtools,xcolor}
\usepackage{hyperref}

\newcommand{\ket}[1]{|#1\rangle}

\begin{document}

\title{Destruction of string order after a quantum quench}

\author{Marcello Calvanese Strinati}
\affiliation{NEST, Scuola Normale Superiore and Istituto Nanoscienze-CNR, I-56126 Pisa, Italy}

\author{Leonardo Mazza}
\affiliation{D\'epartement de Physique, Ecole Normale Sup\'erieure / PSL Research University,
CNRS, 24 rue Lhomond, F-75005 Paris, France}
\affiliation{NEST, Scuola Normale Superiore and Istituto Nanoscienze-CNR, I-56126 Pisa, Italy}

\author{Manuel Endres}
\affiliation{Institute for Quantum Information and Matter, Department of Physics, California Institute of Technology, Pasadena, CA 91125, USA}

\author{Davide Rossini}
\affiliation{NEST, Scuola Normale Superiore and Istituto Nanoscienze-CNR, I-56126 Pisa, Italy}

\author{Rosario Fazio}
\affiliation{ICTP, Strada Costiera 11, 34151 Trieste, Italy}
\affiliation{NEST, Scuola Normale Superiore and Istituto Nanoscienze-CNR, I-56126 Pisa, Italy}

\date{\today}


\begin{abstract}
We investigate the evolution of string order in a spin-1 chain following a quantum quench. After initializing the chain in the Affleck-Kennedy-Lieb-Tasaki state, we analyze in detail how string order evolves as a function of time at different length scales. The Hamiltonian after the quench is chosen either to preserve or to suddenly break the symmetry which ensures the presence of string order. Depending on which of these two situations arises, string order is either preserved or lost even at infinitesimal times in the thermodynamic limit. The fact that non-local order may be abruptly destroyed, what we call \emph{string-order melting}, makes it qualitatively different from typical order parameters in the manner of Landau. This situation is thoroughly characterized by means of numerical simulations based on matrix product states algorithms and analytical studies based on a short-time expansion for several simplified models.
\end{abstract}

\maketitle


\section{Introduction}

Within Landau theory, phases of matter are identified by their local order parameters, and transitions taking place between different phases are successfully understood by analyzing how order parameters change on crossing the transition points. In this framework, local quantities (including two-point correlation functions) fully characterize the phase. Nonetheless, phases that cannot be detected through local quantities do exist. They are usually referred to as topological phases, both protected and non-protected by symmetries~\cite{nayak,wen}. The simplest example of non-local order in one dimension is probably the string order (SO), first discussed in the context of the Haldane phase~\cite{fdmhaldane1,fdmhaldane2}, which characterizes the ground state of integer-spin Heisenberg chains. SO accounts for correlations between spins that are not detected by a simple two-point correlation function~\cite{nijs}, and reveals the presence of a hidden order parameter, that is related to a non-local symmetry of the Hamiltonian~\cite{pollman1,pollman2,kennedy}.

Whereas the resilience of SO to local static perturbations has been thoroughly characterized~\cite{pollman1,pollman2}, much less is known on its robustness to dynamical perturbations. Building on the results reported in an earlier work~\cite{rossinimazzafazioprb}, in this paper we continue the characterization of the time evolution of SO in a specific dynamical protocol: a quantum quench~\cite{polkovnikov}. 
Here, an initially-prepared many-body state evolves under the action of a time-independent Hamiltonian. When the quenched Hamiltonian is extensive, as it is in the case considered in this work, one refers to a global quench. 

The key result upon which this paper is built is the following: SO can be abruptly destroyed after an infinitesimal amount of time. We call this phenomenon \emph{string-order melting}~\cite{rossinimazzafazioprb}. This is in stark contrast with Landau local order, e.g. the magnetization of a spin lattice, where the order parameter of the system cannot disappear instantaneously, but only after a finite amount of time, because of continuity of its time evolution. The goal of this paper is to elucidate the nature of this peculiar situation with numerical simulations and analytical discussions of specific limits.

A number of theoretical works dealt with the problem of studying the dynamics in topological systems after a quantum quench~\cite{tsomokos, hamma2, hamma3, kells, sondhi, caio, rigol, russomanno, privitera}. A general analysis of the behavior of all the indicators of a symmetry-protected topological phase in the presence of a quantum quench is still lacking.  
To the best of our knowledge, the sudden disappearance of SO, one of the mentioned indicators, has not been observed yet, neither theoretically nor experimentally.
Our result opens the intriguing perspective that symmetry-protected topological phases might disappear abruptly in the presence of a time-dependent perturbation. This should be the object of further investigations.

In this paper, we focus on the dynamics of SO in a {spin-1} chain after a global quantum quench. We consider the protocol already introduced in Ref.~\cite{rossinimazzafazioprb}. Initially, the system is 
prepared in the ground state of the Affleck-Kennedy-Lieb-Tasaki (AKLT) Hamiltonian~\cite{aklt1,aklt2}, known as AKLT state, which is the simplest example of a state in the Haldane phase, featuring its typical properties. The persistence of SO after the quench is related to the symmetries of the  Hamiltonian ruling the time evolution~\cite{vestraete}. The case in which the symmetries allow the presence of SO at finite times has been characterized in Ref.~\cite{rossinimazzafazioprb}. Here, we focus on cases in which the symmetries of the quenched Hamiltonian are such that SO is either preserved or suddenly destroyed after the global quantum quench, and characterize this situation in detail.

Our numerical results are based on simulations using matrix product states (MPS)~\cite{uscholwock}. A rich phenomenology can be observed at different length scales. We observe the clear presence of a short-length region, roughly identified as the thermalization region, where correlations induced by the quench have propagated, and of a long-length region, where the physics 
of melting is occurring. We corroborate these results with a simple analytical model which explicitly illustrates the mechanism of SO melting. Properly tuned symmetry properties simplify the analytic description, thus allowing a quantitative description of the dynamics of SO after the quench. 

Understanding the robustness of topological order to a dynamical perturbation is not the only motivation of this work. 
The non-equilibrium dynamics of quantum many-body systems has been theoretically studied since several decades~\cite{niemeijer,barouch}.  
The impressive experimental progresses in the field of ultra-cold atomic gases (see, e.g., \cite{greiner,stroferle,fertig,kinoshita1,gring}) 
spurred a great deal of interest and a consequent renewed theoretical activity~\cite{polkovnikov,eisert}. A prominent portion of this activity on 
non-equilibrium dynamics has  focused on quenches, this special attention being in part motivated by the fact that current experiments can easily implement this protocol.
Thanks to the amazing progresses in the quantum simulation of effective spin models through one-dimensional chains of trapped 
ions and cold atoms~\cite{trotzky,cheneau,monroe1,monroe2}, these questions are not a mere academic curiosity, but retain an experimental 
relevance. In particular, since both string operators~\cite{endres,endresmazza} and the short-time dynamics of closed quantum systems are experimentally 
accessible, it is possible to envision an experimental verification of our results in the near term future.

The paper is organized as follows. In Sec.~\ref{sec:quantumquenchesandstringorder}, we review the main properties of the  Haldane 
phase and of SO in a spin-1 chain. 
In Sec.~\ref{sec:nonlocalorderunderdynamicalperturbations}, we summarize the recent results on out-of-equilibrium SO and report our numerical data on the case of a quantum quench with sudden melting of SO. 
In Sec.~\ref{sec:analyticalresults}, we present an analytical model based on perturbation theory to describe numerical results, and in 
Sec.~\ref{sec:pureakltmodel}, we discuss a simplified model to test the previous analytical results. Conclusions are drawn in Sec.~\ref{sec:conclusions}, while some technical details are provided in the Appendixes. Throughout the paper, we set $\hbar=1$, $k_B=1$.


\section{Haldane phase in the AKLT model}
\label{sec:quantumquenchesandstringorder}

Let us review the basic notions of the Haldane phase that will be useful in the analysis of our results~\cite{pollman1,pollman2}. 
We consider a one-dimensional spin-1 chain governed by the AKLT Hamiltonian~\cite{aklt1,aklt2}
\begin{equation}
\hat{\mathcal H}_{\rm AKLT}
= \sum_n \left[ 
\hat{\mathbf{S}}_n \cdot \hat{\mathbf{S}}_{n+1} 
+ \frac{1}{3} \left( \hat{\mathbf{S}}_n 
\cdot
\hat{\mathbf{S}}_{n+1}  \right)^2
\right] ,
\label{aklthamiltonian}
\end{equation}
where $\hat {\mathbf S}_n = \big(\hat S^x_n, \hat S^y_n, \hat S^z_n \big)^T$ and $\hat S^\alpha_n$ are spin-1 operators.
The ground state $\ket{\Psi_{\rm AKLT}}$, usually known as AKLT state, is exactly known and has a simple MPS representation with bond link $D=2$~\cite{klumper}. It is the simplest example of a state which belongs to the so-called Haldane phase. The presence of non-local order is identified by analyzing the properties of the string operator, defined as
\begin{equation}
\hat{\mathcal{O}}^{(\alpha)}_l \coloneqq
\hat S_k^{\alpha}
\left[\,
\prod_{n=k+1}^{k+l-1}
e^{i\pi \hat{S}_n^{\alpha}}
\right]
\hat S_{k+l}^{\alpha} \, , \quad (\alpha=x,y,z) \,.
\label{definitionofstringorder}
\end{equation}
The system displays SO when:
\begin{equation}
\lim_{l \to \infty} \left\langle \hat{\mathcal{O}}^{(\alpha)}_l \right\rangle 
\neq 0 \,\,, \quad \forall \alpha\,\, .
\label{eq:string:order}
\end{equation}

When open boundary conditions (OBC) are considered, the model defined by $\hat {\mathcal H}_{\rm AKLT}$ exhibits zero-energy spin-$1/2$ 
edge modes. The presence of these edge modes makes the AKLT state four-fold degenerate on an open chain, whereas in the presence of periodic boundary conditions it is non-degenerate. 

This phenomenology is intimately related to the existence of SO. Indeed, both these features can be regarded as a consequence of a 
hidden symmetry breaking related to the $\mathbb D_2$ group:
\begin{equation}  
  \mathcal G_{\mathbb D_2} \coloneqq \left\{ 
  \hat {\mathbb I},
  e^{-i \pi \sum_n \hat S_n^x},
  e^{-i \pi \sum_n \hat S_n^y},
  e^{-i \pi \sum_n \hat S_n^z} 
  \right\} \,\, .
\end{equation}
Note that even if $\hat {\mathcal H}_{\rm AKLT}$ is left invariant by the action of $\mathcal G_{\mathbb  D_2}$, there is no contradiction in the fact that symmetry-breaking might occur in a one-dimensional system at zero-temperature because the group is finite. The AKLT open chain can be mapped onto a local ferromagnetic chain with four symmetry-broken magnetic states by means of a non-local transformation (the Kennedy-Tasaki transformation)~\cite{kennedy}. String operators in the AKLT chain are mapped onto two-point spin correlators which reveal ferromagnetic order, and thus, in this dual picture, string operators reveal the presence of hidden, long-range antiferromagnetic order, as earlier proposed in Ref.~\cite{nijs}. SO can be considered as the nonlocal order parameter which identifies states in the Haldane phase, and its relation to hidden long-range order has also been discussed in different contexts~\cite{egdallatorre1,egdallatorre2,hasebe,chhajlany}.

Because of the special role of the symmetry group $\mathcal G_{\mathbb D_2}$, the Haldane phase is better referred to as a symmetry-protected topological phase. Indeed, it has been shown that a perturbation $\hat V$ which is not left invariant by the group $\mathcal G_{\mathbb D_2}$ destroys SO in the ground state. More rigorously, it has been demonstrated that the presence of SO is equivalent to the existence of a local symmetry~\cite{cirac}; in the case of the AKLT state, this means that SO is related to the fact that $\ket{\Psi_{\rm AKLT}}$ is invariant with respect to the action of $\mathcal G_{\mathbb D_2}$.
In the absence of such symmetry, SO will exponentially vanish on increasing the length of the string~\cite{cirac}.

\section{Non-local order under dynamical perturbations}
\label{sec:nonlocalorderunderdynamicalperturbations}
\subsection{Quantum quenches and string-order melting}
\label{sec:stringorderandquantumquenches}

In this section, we start recapitulating the theory of global quantum quenches applied to SO developed in Ref.~\cite{rossinimazzafazioprb}. We then define the concept of SO melting and present some general considerations on it.

The quench protocol is defined as follows. Consider a system at zero temperature
described by the Hamiltonian $\hat{\mathcal H}(\kappa)$ which depends on a parameter $\kappa$. Let $\kappa_0$ be the initial value of $\kappa$, so that at the beginning the system is in the ground state of $\hat{\mathcal H}(\kappa_0)\equiv\hat{\mathcal{H}}_0$. 
At time $t=0$, $\kappa$ is abruptly varied from $\kappa_0$ to a new value $\kappa_1\neq \kappa_0$ such that the new Hamiltonian is 
$\hat{\mathcal{H}}(\kappa_1)\equiv\hat{\mathcal{H}}$, and thus the system is suddenly driven out of equilibrium. If we denote with $|\Psi_0\rangle$ the ground 
state of $\hat{\mathcal H}_0$, the state at time $t$ is $|\Psi(t)\rangle=e^{-i \hat{\mathcal H}t}|\Psi_0\rangle$. 
As mentioned in the Introduction, we consider $|\Psi_0\rangle\equiv\ket{\Psi_{\rm AKLT}}$. 

The discussion of SO at finite times is intimately related to the symmetry properties of 
$\ket{\Psi_{\rm AKLT}}$ and $\hat {\mathcal H}$; we thus recall some notions of group theory. Let us begin by stressing that $\mathcal{G}_{\mathbb{D}_2}$ is an Abelian group and admits four irreducible (one-dimensional) representations: $\Gamma_0$ (the identity), $\Gamma_x$, $\Gamma_y$ and $\Gamma_z$.
From now on, we dub \emph{even} an operator $\hat A_e$ which transforms under action of the group according to the representation 
$\Gamma_0$, i.e. it is left invariant $\hat g \hat A_e \hat g^\dagger = \hat A_e$, $\forall\hat g\in\mathcal{G}_{\mathbb{D}_2}$. An operator $\hat A_o$ is 
dubbed \emph{odd} if, on the contrary, it is a linear combination of operators which transform under the action of the group according to the representations 
$\Gamma_x$, $\Gamma_y$ and $\Gamma_z$. The Hamiltonian $\hat {\mathcal H} $ can be formally decomposed as $\hat {\mathcal H}_e + 
\hat {\mathcal H}_o $.

The necessary and sufficient condition for the simultaneous existence of SO at infinitesimally short times along all the three
directions $\alpha = x,y,z$ is that 
\begin{equation}
 \hat {\mathcal H}_o
 \big( \hat {\mathcal H_e} \big)^n
 \ket{\Psi_{\rm AKLT}} = 0\,\, , \quad 
 \forall n \in \mathbb N \,\, .
 \label{eq:check}
\end{equation}
It is very important to stress that this result depends on the fact that we explicitly consider a state which displays SO at time $t=0$.
If Eq.~\eqref{eq:check} is not satisfied, SO is not present for $t>0$, and thus it disappears abruptly for at least one of the axes $x$, $y$, or $z$. If Eq.~\eqref{eq:check} is fulfilled, the time evolution of SO is continuous and SO is present not only at infinitesimally short times, but up to a finite time $\tau$.
This result was first discussed in Ref.~\cite{rossinimazzafazioprb}, to which the interested reader is referred. For completeness, we report the proof of Eq.~\eqref{eq:check} in Appendix~\ref{appendixproofofequation}.

The previous relation highlights the role of symmetry: if $\hat {\mathcal H}_o=0$, the 
evolved state $|\Psi(t)\rangle$ has the same symmetry of $|\Psi_{\rm AKLT}\rangle$ at any time, and SO is well defined at finite short times.
If $\hat {\mathcal H}_o \neq 0$, two situations may arise. 
If the relation~\eqref{eq:check} is satisfied, the state nevertheless keeps the original symmetry, and SO is well defined.
However, when Eq.~\eqref{eq:check} is not fulfilled, SO is abruptly lost. 
This is possible because, although the dynamics of every string operator $\hat {\mathcal O}^{(\alpha)}_{l}$ is continuous in time, the limit procedure contained in Eq.~\eqref{eq:string:order} introduces a singularity in the thermodynamic limit. As we will see, the expectation value of the string operator decays as the string length increases, approaching the zero value in the thermodynamic limit. This is the situation which, in this paper, is referred to as SO melting.

\subsection{Numerical results}
\label{sec:numericalresults}

\begin{figure*}[t]
\centering
\includegraphics[width=4.9cm]{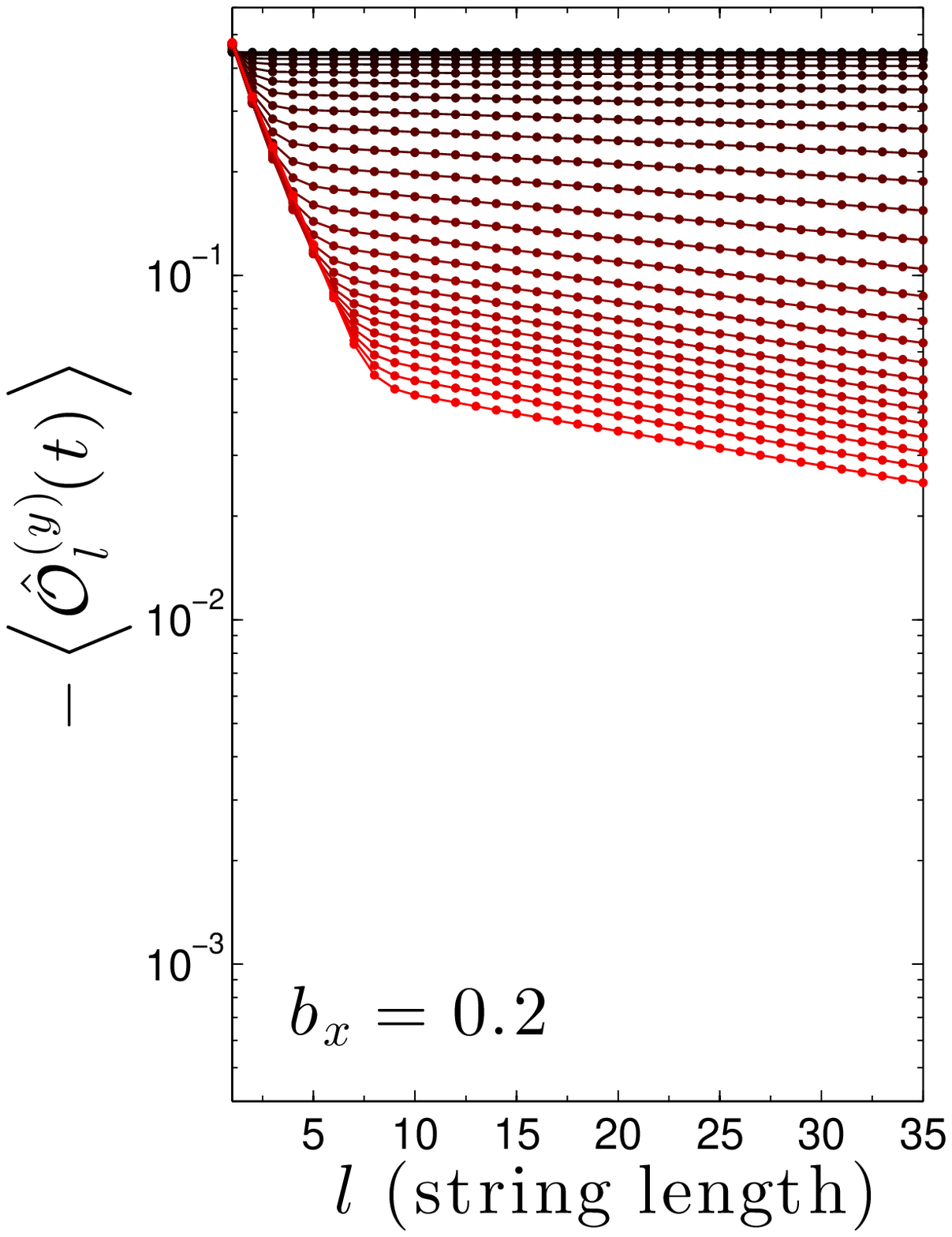}
\hspace{-0.3cm}
\includegraphics[width=4.3cm]{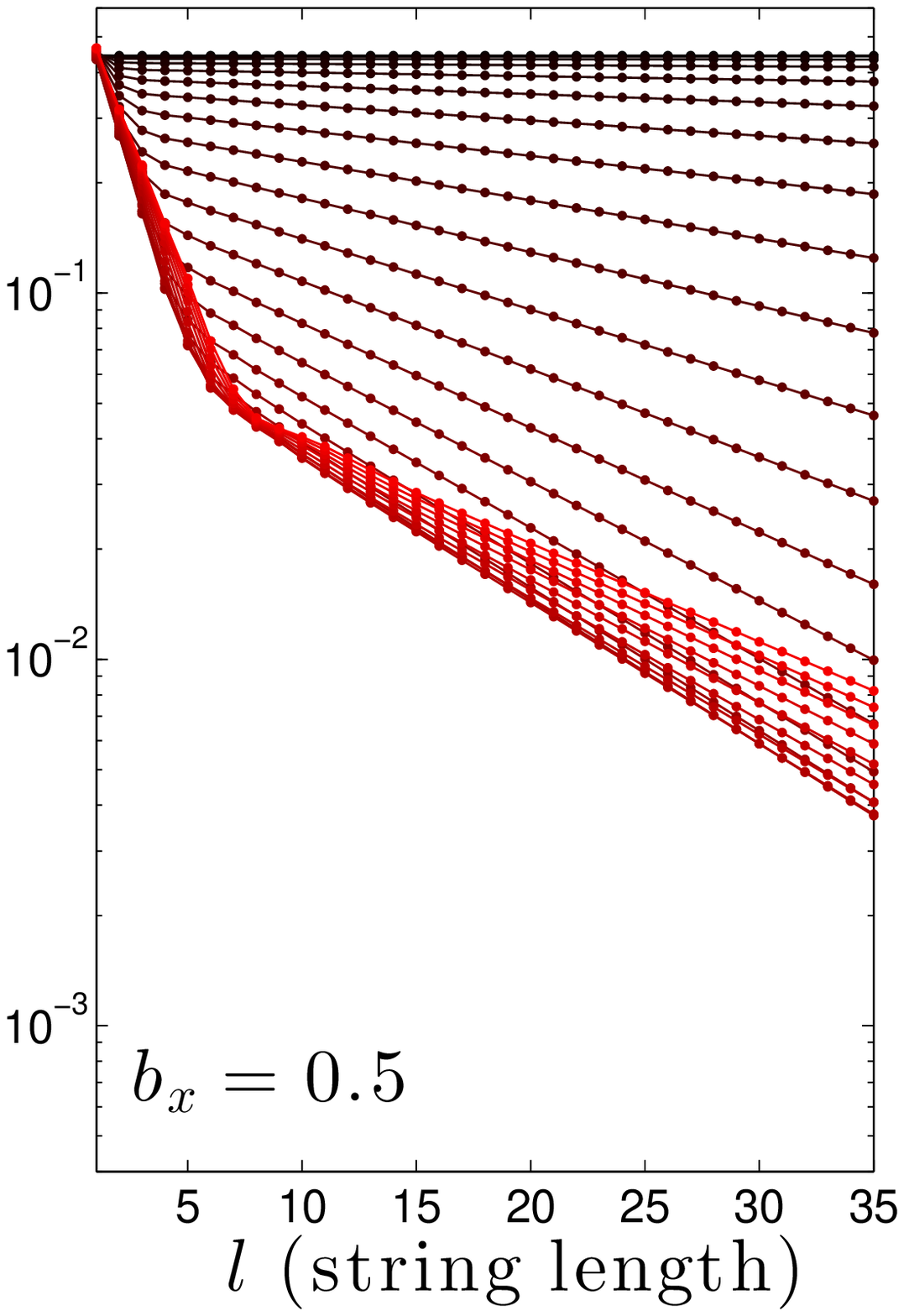}
\hspace{-0.3cm}
\includegraphics[width=4.3cm]{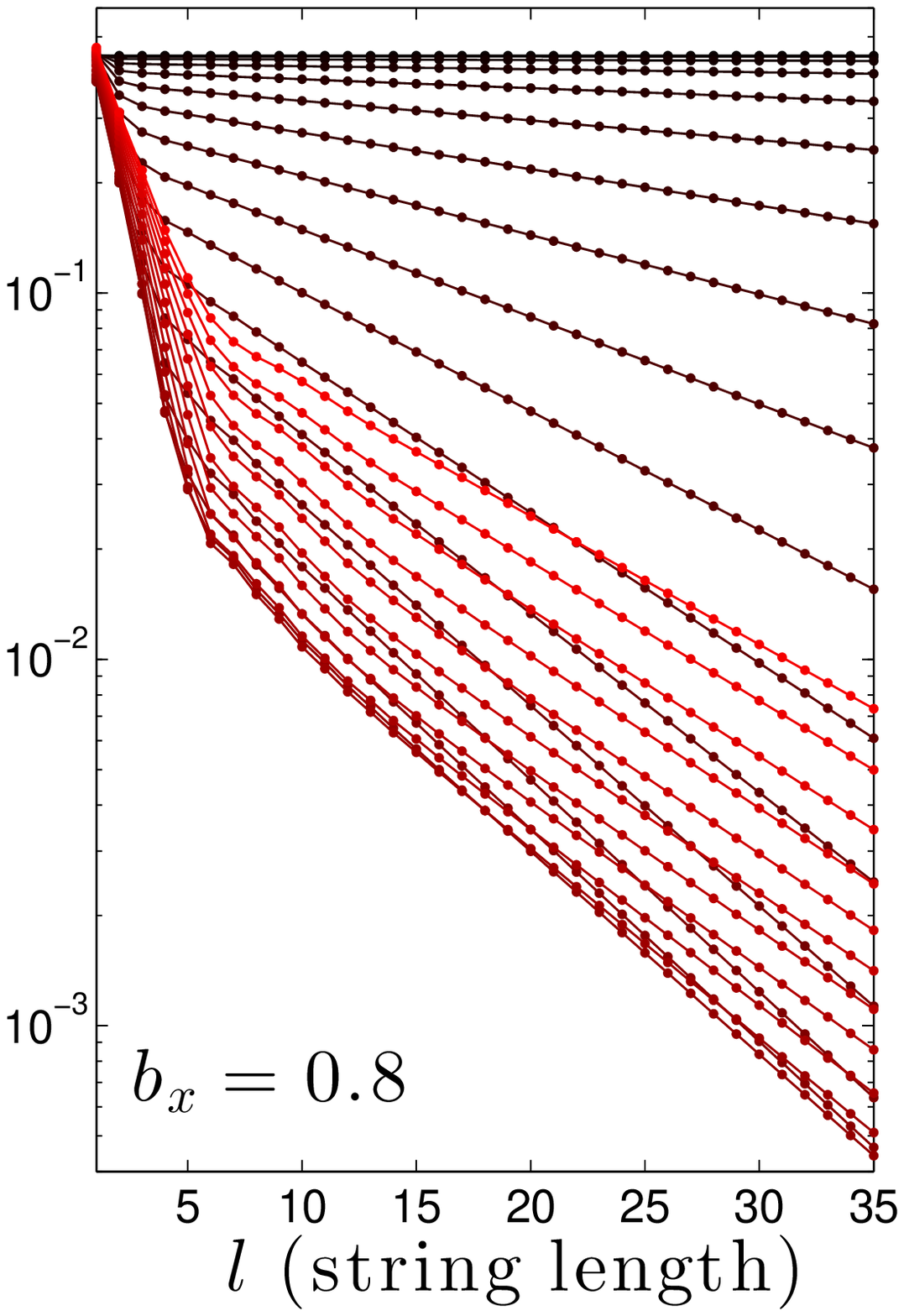}
\hspace{-0.3cm}
\includegraphics[width=4.3cm]{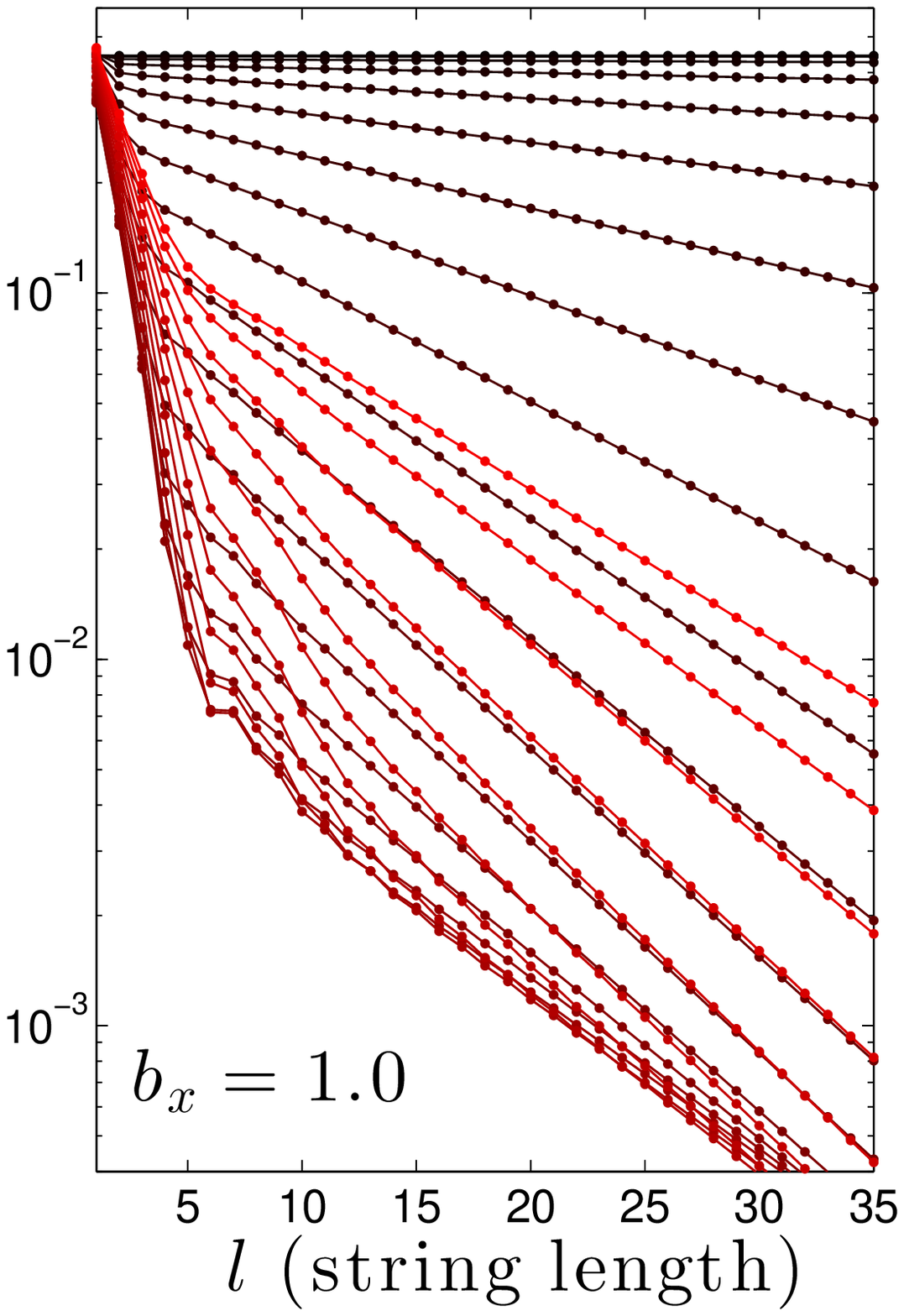}
\includegraphics[width=4.9cm]{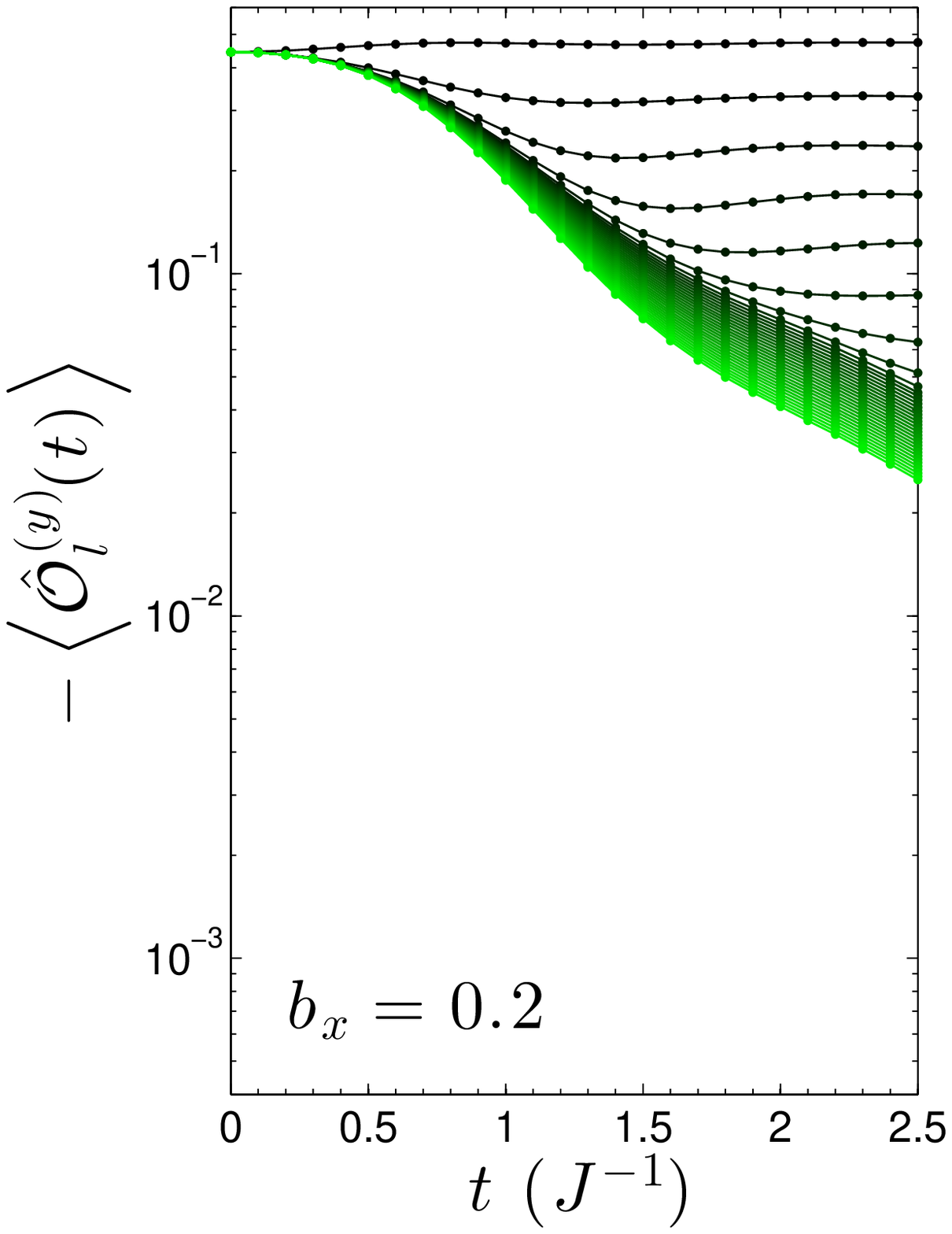}
\hspace{-0.3cm}
\includegraphics[width=4.3cm]{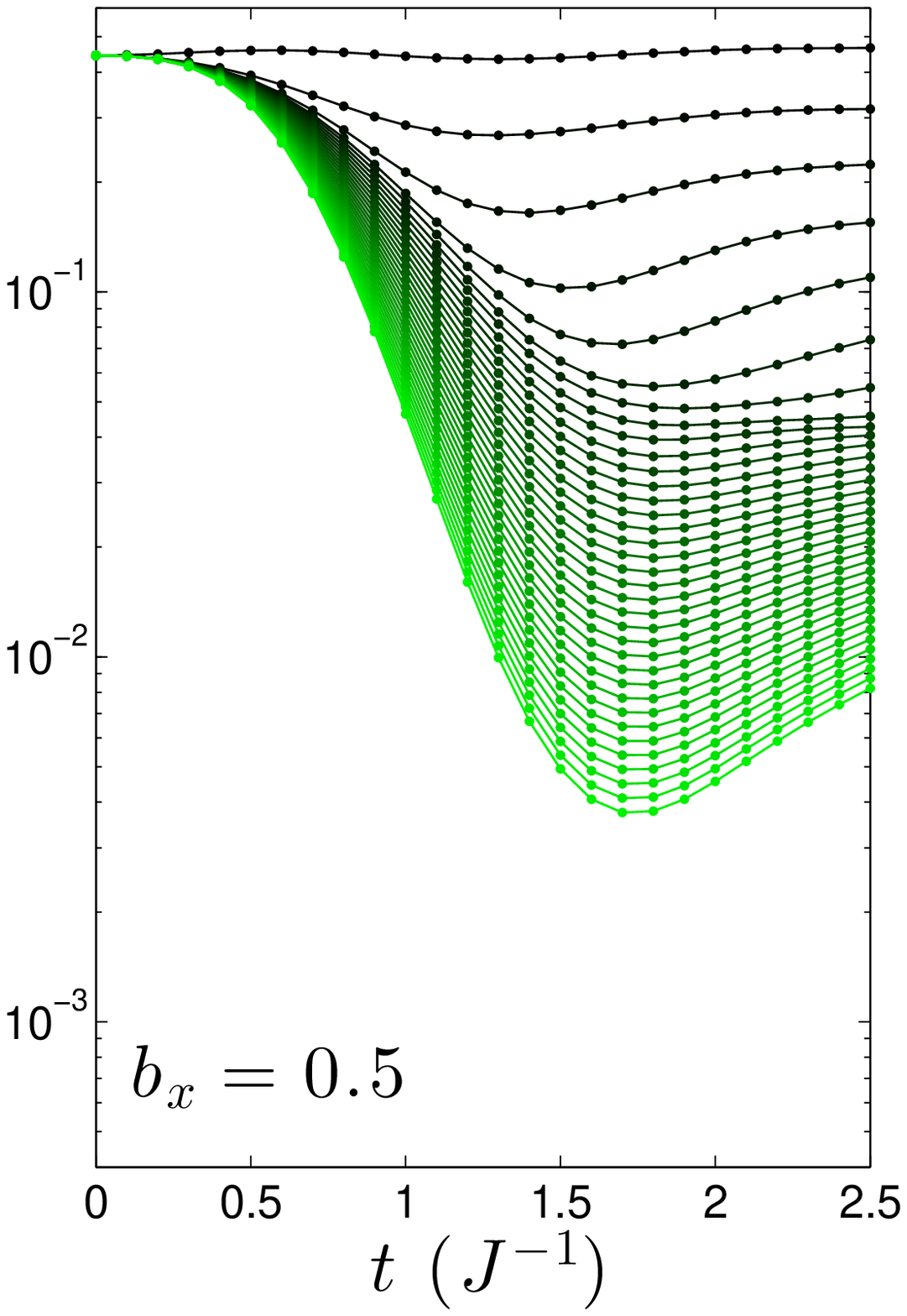}
\hspace{-0.3cm}
\includegraphics[width=4.3cm]{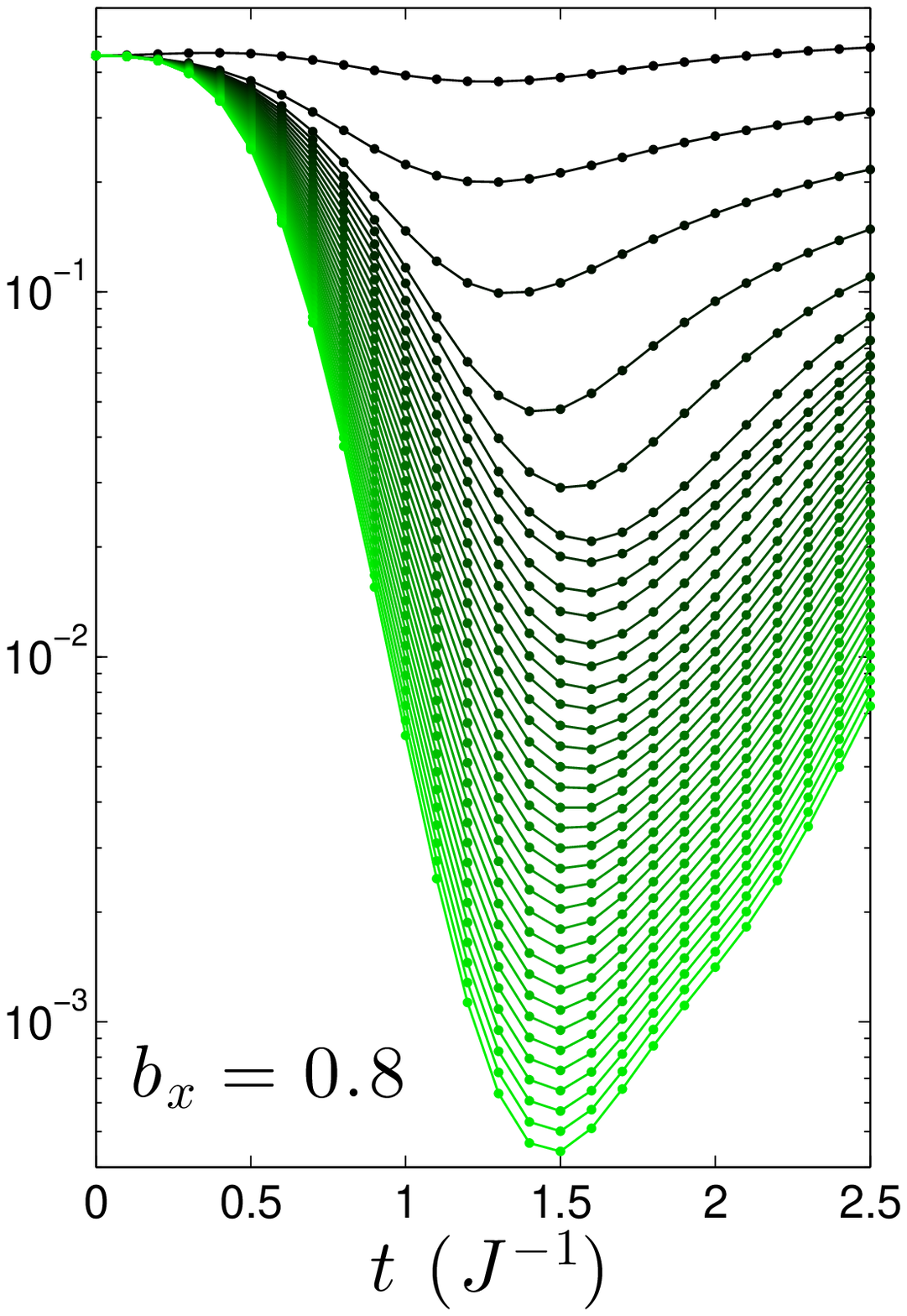}
\hspace{-0.3cm}
\includegraphics[width=4.3cm]{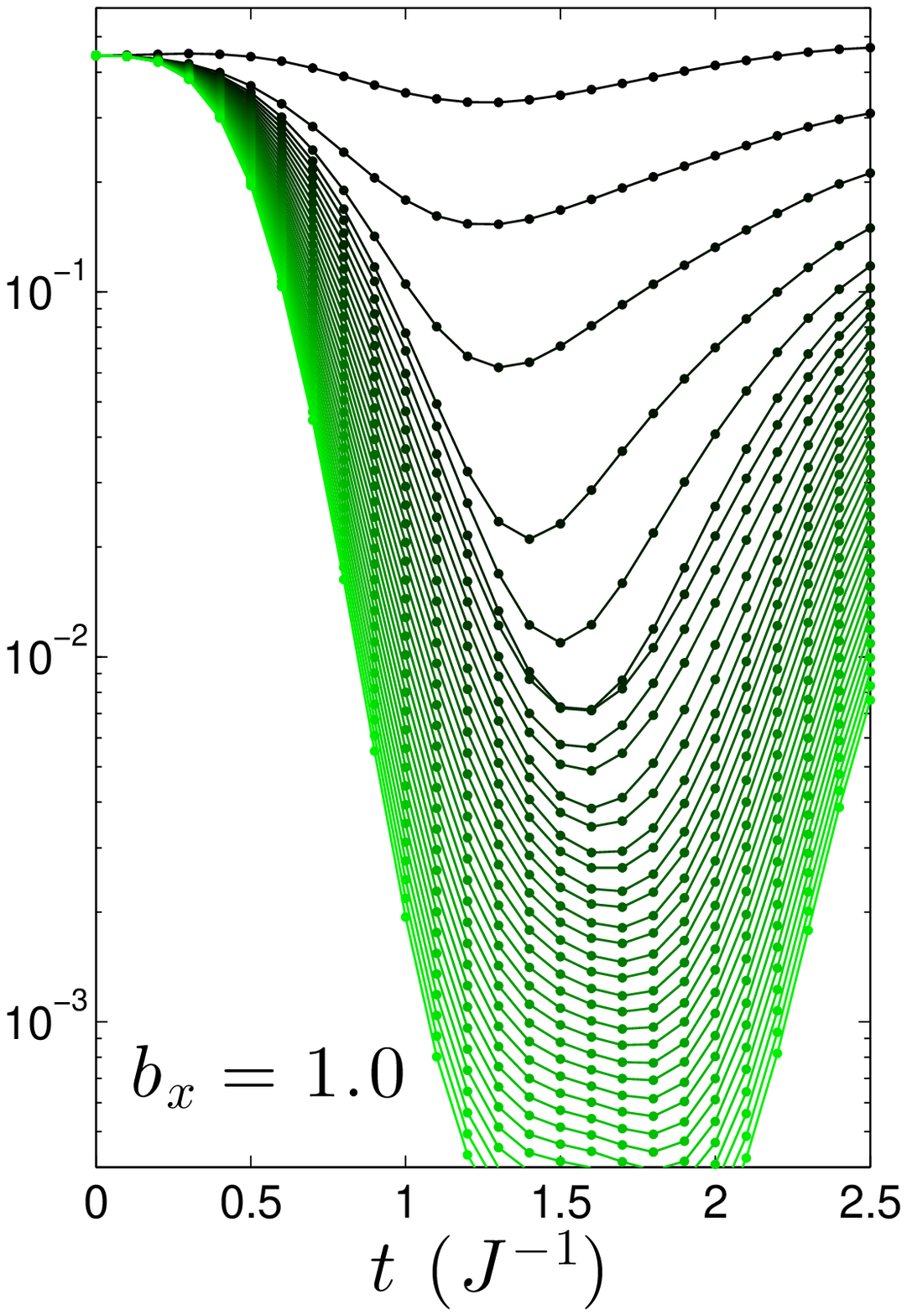}
\caption{(Color online) Results of the numerical simulations described in Sec.~\ref{sec:numericalresults}. We simulate $L=60$ sites with OBC. 
(Upper panels) Expectation value of the string operator along the $y$-axis as a function of the string 
length $l$ and at fixed time after a quench using the Hamiltonian in Eq.~\eqref{xxzhamiltonian} with $\Delta=0.2$ and odd perturbation $\sum_n \hat S_n^x$ for four different values of the coupling constant, here indicated with $b_x$. 
Each line corresponds to a different time, from $t=0$ (dark red) to $t=2.5\,J^{-1}$ (bright red) with time spacing $0.1\,J^{-1}$. For short lengths a steady region appears, whose size increases with time, and for longer lengths SO melting in the bulk is observed. 
(Lower panels) Expectation value of the string operator along the $y$-axis as a function of time at fixed string length, from $l=1$ (dark green) to $l=35$ (bright green). 
For $b_x=0.2$ the expectation value of the string operator becomes independent of time (a local stationary state is reached) at short lengths. 
For larger values of $b_x$, simulations are too short to reveal a steady region at short lengths. Thermalization of SO at 
short lengths to the corresponding canonical ensemble with effective temperature $\beta\simeq 1.5\,J^{-1}$ is reported in Fig.~\ref{fig:thermalizationofstringorderxxz} for $b_x=0.2$.}
\label{fig:oldsimulationsxxzodd}
\end{figure*}

We now present numerical simulations of a quantum quench where SO melting takes place.
After initializing a spin-1 one-dimensional chain in the AKLT state $\ket{\Psi_{\rm AKLT}}$, we perform a quantum quench with the Hamiltonian
\begin{equation}
\hat{\mathcal H} =
J \sum_i \left[
\hat S^{x}_i \hat S^{x}_{i+1} +
\hat S^{y}_i \hat S^{y}_{i+1} +
\Delta \hat S^{z}_i \hat S^{z}_{i+1} +
b_x \hat S^x_i
\right] \,\, .
\label{xxzhamiltonian}
\end{equation}
When $b_x = 0$, we recover the XXZ Hamiltonian $\hat{\mathcal H}_{\rm XXZ}$ which identifies the even part $\hat{\mathcal{H}}_e$. The spin-1 XXZ model  is known to be in the Haldane phase for $\Delta<\Delta_c=1.186...$~\cite{boschi,hueda}. Since it is $\mathbb{D}_2$-invariant, no SO melting is induced; this is the case which has been studied in Ref.~\cite{rossinimazzafazioprb}. In Eq.~\eqref{xxzhamiltonian}, the odd part of the Hamiltonian is given by $\hat{\mathcal{H}}_o=b_x\sum_i\hat S^x_i$.

The numerical procedure that we employ here consists of two steps: 
\begin{itemize}

\item Even if the AKLT state has an exact MPS representation, for technical reasons we find the MPS representation of $\ket{\Psi_{\rm AKLT}}$ by means of a variational search in the MPS space~\cite{uscholwock} using the AKLT Hamiltonian. We start from a random MPS state with initial bond link $D=10$ and iterate the variational procedure sweeping the chain until convergence is reached, i.e. the ground-state energy reaches a constant value. In our simulations we adopt OBC.

\item After the ground state is found, we quench the system with the XXZ Hamiltonian (\ref{xxzhamiltonian}) with $\Delta = 0.2$ and compute the time-evolved state $|\Psi(t)\rangle$ by means of time-evolving block decimation (TEBD) algorithm~\cite{uscholwock,vidal1,vidal2}. The time evolution is decomposed by using a fourth-order Trotter expansion~\cite{suzuki1,suzuki2}. Finite-temperature simulations are carried out by using the purification method of a mixed quantum state~\cite{uscholwock,vestraete} using a sixth-order Trotter expansion.
\end{itemize}

In order to observe SO melting, $b_x \neq 0$ is a necessary condition. The results are shown in Fig.~\ref{fig:oldsimulationsxxzodd}, where we consider a chain of length $L=60$. We plot the expectation value of the string operator along the $y$-axis $\left\langle \hat{\mathcal{O}}^{(y)}_l \right\rangle(t)$ for $b_x=0.2,0.5,0.8,1.0$. 
Because of the choice of OBC, to compute the expectation value of the string operator we have to discard those sites close to the chain ends; we choose to discard the $10\%$ of the chain from both ends.

The results displayed in Fig.~\ref{fig:oldsimulationsxxzodd}, top panels, show that $\left\langle \hat{\mathcal{O}}^{(y)}_l \right\rangle(t)$ depends exponentially on $l$, the string length. However, there are two typical decay lengths, one which is valid for short length scales, and one which is valid for long length scales. 
We first address the physics of the system at short length scales. The time dependence of $\left\langle \hat{\mathcal{O}}^{(y)}_l \right\rangle(t)$ (Fig.~\ref{fig:oldsimulationsxxzodd}, bottom panels) shows a steady behavior for short string lengths.

\begin{figure}
\centering
\includegraphics[width=8.6cm]{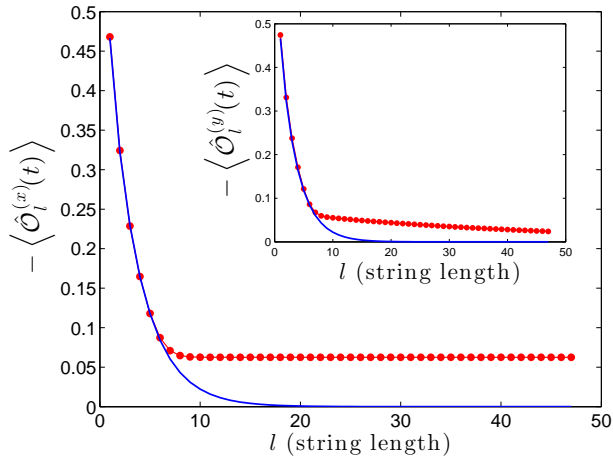}
\caption{(Color online) Thermalization analysis of the data in Fig.~\ref{fig:oldsimulationsxxzodd}. The system is quenched with Hamiltonian 
(\ref{xxzhamiltonian}) by using $\Delta=0.2$ and $b_x=0.2$ (see Fig.~\ref{fig:oldsimulationsxxzodd}, leftmost panels). The thermal average 
of the expectation value of the string operator along the $x$ axis as a function of string length (blue thick line) with inverse temperature 
$\beta \simeq 1.5\,J^{-1}$ is overlapped to the expectation value of the string operator along the $x$ axis as a function of the string length, computed at fixed time $t=2.28\,J^{-1}$ (red dots). In the inset we show the same analysis for the expectation value of the string operator along the 
$y$ axis. In both cases we numerically observe thermalization of SO at short lengths.}
\label{fig:thermalizationofstringorderxxz}
\end{figure}

We now provide evidence that this short-length behavior is related to the occurrence of thermalization. We show in Fig.~\ref{fig:thermalizationofstringorderxxz} 
the expectation value of the string operator along the $x$ and $y$ axes as a function of string length at fixed time $t=2.28\,J^{-1}$.  The numerical 
data are compared with the thermal average $\mathcal{Z}^{-1}{\rm Tr} [\hat{\mathcal{O}}_l^{(\alpha)}\,e^{-\beta\hat{\mathcal{H}}}]$ for 
$\alpha=x,y$, where $\mathcal{Z}$ is the corresponding partition function, and $\hat{\mathcal{H}}$ is given by Hamiltonian 
(\ref{xxzhamiltonian}) with $\Delta=0.2$ and $b_x=0.2$. The inverse temperature $\beta$ is defined, as customary,  by the relation 
$\langle\Psi_0|\hat{\mathcal{H}}|\Psi_0\rangle=\mathcal{Z}^{-1}{\rm Tr} [\hat{\mathcal{H}}\,e^{-\beta\hat{\mathcal{H}}} ]$. From our simulations, by matching the energy of the AKLT state with that of a thermal state, we find $\beta \simeq 1.5\,J^{-1}$. As shown in Fig.~\ref{fig:thermalizationofstringorderxxz}, the properties of this thermal state describe with very good fidelity the long-time and short-length behavior of our system, thus pointing out the occurrence of thermalization in this model. This result complements the occurrence of thermalization discussed at $b_x=0$ in Ref.~\cite{rossinimazzafazioprb}.

We now move to the discussion of the system at long length scales. From Fig.~\ref{fig:oldsimulationsxxzodd} we observe that by increasing the strength of $b_x$, the long-length decay becomes more pronounced, without significantly affecting the thermalized region. This effect is a clear indication of the radically different physical origin of  the two distinct exponential decays. The long-length regime is the one associated to the melting of SO.

In order to better understand the melting process and its interplay with thermalization, we now proceed with a combined analytical and numerical analysis. 
The problem of describing the expectation value of the string operator as a function of time and of string length can be tackled by means of perturbation theory. By considering the strength of the magnetic field as a small expansion parameter, it is possible to explicitly see the relation between the symmetry properties of the post-quench Hamiltonian and SO. 
This is the subject of the following sections.

Before concluding, let us comment on the fact, displayed in Fig.~\ref{fig:thermalizationofstringorderxxz}, that SO along the $x$ axis survives the quench. This is not violating the results in Sec.~\ref{sec:stringorderandquantumquenches}, which refer to the symmetry properties of the Hamiltonian which are necessary and sufficient to ensure the existence of SO along all the three axes $\alpha = x,y,z$. Here, SO melts along $y$ and $z$ axes (not shown) and thus there is no contradiction. In Sec.~\ref{sec:quenchwiththexxzhamiltonianinatransversefield} we discuss the symmetry properties of a quench Hamiltonian in order to let SO survive only along some of the three Cartesian axes.

Finally, we stress that when the system is quenched with a magnetic field along the $z$ axis, the presence of the magnetic field has no effect on the dynamics of SO. This follows from the fact that  $\left[\hat{\mathcal{H}}_{\rm XXZ},\sum_n \hat{S}_n^z\right]=0$ and $\sum_n \hat{S}_n^z|\Psi_{\rm AKLT}\rangle=0$.
By using $\hat{\mathcal{H}}_e\equiv\hat{\mathcal{H}}_{\rm XXZ}$ and $\hat{\mathcal{H}}_o\equiv\sum_n\hat{S}_n^z$, it is straightforward to see that Eq.~\eqref{eq:check} is fulfilled. 

\subsection{Rotating string order}

A curious question is whether, in the presence of SO melting, there exists a \emph{rotating} frame of reference in which SO is restored. Consider the case of a \emph{rotating string operator}, defined by
\begin{equation}
\hat{\mathcal{R}}^{(\alpha)}_l(t)\coloneqq \hat{U}^\dag_r(t)\,\hat{\mathcal{O}}^{(\alpha)}_l\,\hat{U}_r(t) \,\, .
\label{stringorderrotaingoperator}
\end{equation}
where the \emph{rotating operator} is $\hat{U}_r(t)=e^{i\hat{\mathcal{H}}'_ot}$. We take an odd Hamiltonian given by the sum of only on-site operators, i.e., $\hat{\mathcal{H}}'_o=\sum_j\hat{h}'_{o,j}$, where $\hat{h}'_{o,j}$ has support on the $j$th site only; this requirement is crucial in order to preserve the local structure of $\hat{\mathcal{O}}^{(\alpha)}_l$ when acting with $\hat{U}_r(t)$. Equation~\eqref{stringorderrotaingoperator} becomes
\begin{equation}
\hat{\mathcal{R}}^{(\alpha)}_l(t)=\hat S^{\alpha}_k(t)\left[\,\prod_{n=k+1}^{k+l-1}e^{i\pi\hat S^{\alpha}_n(t)}\,\right]\hat S^{\alpha}_{k+l}(t) \,\, ,
\label{stringorderrotaingoperatorlocal}
\end{equation}
where $\hat S^{\alpha}_j(t)=e^{-i\hat h_{o,j}'t}\hat S^{\alpha}_je^{i\hat h_{o,j}'t}$. The rotating SO parameter in Eq.~\eqref{stringorderrotaingoperatorlocal} could be measured in experiments by applying a global spin-rotation pulse before the detection.

The introduction of the rotating SO operator~\eqref{stringorderrotaingoperatorlocal} allows us to show under which conditions the rotating operator $\hat{U}_r(t)$ restores SO after the quench.
Let us assume that Eq.~\eqref{eq:check} does not hold,
namely, that there is SO melting.
If we consider for instance the situation: (i) $\hat {\mathcal H}'_o \equiv \hat{\mathcal H}_o$ and (ii) $[\hat{\mathcal{H}}_e,\hat{\mathcal{H}}_o]=0$,
we can show that, at short times, the time-evolved system displays rotating SO:
\begin{equation}
 \lim_{l \to \infty} \langle \Psi(t)| \hat{\mathcal{R}}^{(\alpha)}_l (t) 
 |\Psi(t)\rangle 
\neq 0 \,\,, \quad \forall \alpha\,\, .
\label{eq:rotating:string:order}
\end{equation}

In order to show this statement, we observe that Eq.~\eqref{eq:rotating:string:order} is equivalent to requiring that the state $\hat{U}_r(t)|\Psi(t)\rangle=e^{i\hat{\mathcal{H}}'_ot}\,e^{-i(\hat{\mathcal{H}}_e+\hat{\mathcal{H}}_o)t}|\Psi_0\rangle$ is $\mathbb{D}_2$-invariant. Indeed, an expansion at first order in time yields:
\begin{equation}
 \hat{U}_r(\delta t)|\Psi(\delta t)\rangle\simeq|\Psi_0\rangle-i\,\delta t\,
 \left[  \hat{\mathcal{H}}_e+ \hat{\mathcal{H}}_o- \hat{\mathcal{H}}'_o  \right]
 |\Psi_0\rangle.
 \label{eq:first_order}
\end{equation}
It follows that $\mathbb D_2$ invariance is present only if $(\hat{\mathcal{H}}_o-\hat{\mathcal{H}}'_o)|\Psi_0\rangle=0$. 
This is clearly true in the case 
$\hat{\mathcal{H}}_o=\hat{\mathcal{H}}'_o$
corresponding to hypothesis (i).
Moving to second order, we obtain:
\begin{eqnarray}
\hat{U}_r(\delta t)|\Psi(\delta t)\rangle&\simeq&|\Psi_0\rangle-i\,\delta t\,\hat{\mathcal{H}}_e|\Psi_0\rangle+\nonumber\\
\nonumber\\
&&-\frac{{(\delta t)}^2}{2}\left({\hat{\mathcal{H}}_e}^2+[\hat{\mathcal{H}}_e,\hat{\mathcal{H}}_o]\right)|\Psi_0\rangle+\dots \,\, .\nonumber\\
\label{movedtimeevolvedstate}
\end{eqnarray}
When $[\hat{\mathcal{H}}_e,\hat{\mathcal{H}}_o]=0$, which is hypothesis (ii), the state $\hat{U}_r(t)|\Psi(t)\rangle$ displays SO, since  $\mathbb{D}_2$ symmetry is fully preserved at all orders in $t$. 

Conditions (i) and (ii) are thus sufficient to ensure rotating SO, although they unfortunately select only a limited range of situations. 
A generalization of such conditions would give a more comprehensive understanding of the phenomenon of SO in a time-evolved frame.
In this paper, we do not provide a detailed analytical and numerical analysis of this fact, which remains a subject of further study.

We conclude by showing that, in the case studied numerically in Sec.~\ref{sec:numericalresults}, 
the melting of SO along the $y$ axis is not associated with the existence of SO with respect to a rotating spin given by $\hat S^{y}_j(t)=e^{-i\hat S^{x}_{j}t}\hat S^{y}_j e^{i\hat S^{x}_{j}t}$. 
In this case, 
$$\hat {\mathcal H}_o = \hat {\mathcal H}_o' = \sum_n \hat S^x_n,$$ 
and thus the state $\hat{U}_r(t)|\Psi( t)\rangle$ is $\mathbb D_2$-invariant at first order in $t$.
However, $[\hat{\mathcal{H}}_{e},\hat {\mathcal H}_o]\neq0$, and in particular it can be shown that $[\hat{\mathcal{H}}_{e},\hat {\mathcal H}_o]|\Psi_{\rm AKLT}\rangle\neq0$.
Using Eq.~\eqref{movedtimeevolvedstate}, one observes that the state $\hat{U}_r( t)|\Psi( t)\rangle$ is not $\mathbb D_2$-invariant. This is sufficient to ensure the absence of rotating SO~\eqref{eq:rotating:string:order} at finite times.


\section{Perturbative expansion}
\label{sec:analyticalresults}

A more detailed analysis of the onset of SO melting can be achieved by considering the dynamics of the string at short times. Here 
it is possible to construct a perturbative expansion that we are going to compare with  the numerical simulations.

\subsection{Expectation value in the interaction picture}

In order to perform a short-time expansion, we use the interaction picture to describe the evolution of the state after the quench. The general description is the following: let us consider the even Hamiltonian before the quench $\hat {\mathcal H}_0$, and its even ground state $|\Psi_0\rangle$, such that $\hat g \ket{\Psi_0} = \ket{\Psi_0}$ $\forall\hat g \in \mathcal G_{\mathbb D_2}$. The Hamiltonian which rules the dynamics is decomposed as in Sec~\ref{sec:stringorderandquantumquenches}. Since we are going to develop a perturbative treatment, it is convenient to rename $\hat {\mathcal H}_o = \epsilon \hat V$, where $\| \hat V \|_{\rm op} \sim \| \hat {\mathcal H}_e \|_{\rm op} $ and $\epsilon$ is a dimensionless coupling constant. Here $\| \hat V \|_{\rm op}$ denotes the operatorial norm of the operator $\hat V$ defined by $\| \hat V \|_{\rm op}\coloneqq{\rm sup}_{|\psi\rangle:\langle \psi|\psi\rangle=1}\|\hat V|\psi\rangle\|$. The time-evolved state is $\ket{\Psi (t)} = e^{-i \hat {\mathcal H} t} \ket{\Psi_0}$. Let $\hat A$ be a generic observable, then its time evolution after the quench reads as
\begin{equation}
\langle \hat A \rangle(t)=
\langle\Psi(t)|\hat A|\Psi(t)\rangle \,\, .
\end{equation}
Time evolution can be recast in the interaction picture:
\begin{subequations}
\label{interactionpicture}
\begin{align}
|\Psi (t) \rangle_I & =
e^{i\hat {\mathcal H}_e t}
|\Psi (t)\rangle_S \, ,\\
\hat A_I (t) & =
e^{i \hat{\mathcal H}_et}\,
\hat A_S
\,e^{-i \hat{\mathcal H}_et} \,\, ,
\end{align}
\end{subequations}
where the subscripts $S$ and $I$ denote Schr\"odinger and interaction pictures respectively. In the interaction picture, the wavefunction evolution can be recast in terms of the propagator $\hat U(t)$, such that $|\Psi(t)\rangle_I=\hat U(t)|\Psi(0)\rangle_S$ 
and defined as
\begin{equation}
\hat U(t)=\sum_{n=0}^{\infty}{(-i\epsilon)}^n\int_{0}^{t}d\tau_1\dots\int_{0}^{\tau_{n-1}}d\tau_n\,\hat V_I(\tau_1)\dots\hat V_I(\tau_n) \,\, ,
\label{timeevolutionpropagator}
\end{equation}
where $\hat V_I(t)=e^{i\hat{\mathcal{H}}_et}\,\hat V\,e^{-i\hat{\mathcal{H}}_et}$ is the perturbation in the interaction picture. After Eqs.~\eqref{interactionpicture}, the wave function in the Schr\"odinger picture reads as $|\Psi(t)\rangle_S=e^{-i\hat{\mathcal{H}}_et}\,\hat U(t)|\Psi(0)\rangle_S$. The subscripts $S$ and $I$ will be henceforth omitted, and since at time $t=0$ (before quenching the system) the chain is in the state $|\Psi_0\rangle$, the time-evolved state becomes $|\Psi(t)\rangle=e^{-i\hat{\mathcal{H}}_et}\,\hat U(t)|\Psi_0\rangle$. The time evolution of the operator $\hat A$ is then given by
\begin{equation}
\langle\hat A\rangle(t)=\langle\Psi_0|\hat U^\dag(t)\hat A(t)\hat U(t)|\Psi_0\rangle \label{timeevolutiongenericoperator}\,\, ,
\end{equation}
where $\hat A(t)=e^{i\hat{\mathcal{H}}_et}\,\hat A\,e^{-i\hat{\mathcal{H}}_et}$. The time integrals in Eq.~\eqref{timeevolutionpropagator} are evaluated by 
expanding the operators as
\begin{equation}
\hat V(t)=\sum_{n=0}^{\infty}\frac{{(i)}^{n}}{n!}\,{{\rm ad}({\mathcal{\hat H}_e})}^n(\hat V)\,t^n \label{timeexpansionoperator}\,\, ,
\end{equation}
where we adopt the notation ${{\rm ad}({\mathcal{\hat H}_e})}^n(\hat V)$ for the nested commutator of order $n$ between $\mathcal{\hat H}_e$ and $\hat V$, ${\rm ad}({\mathcal{\hat H}_e})(\hat V)\equiv[\mathcal{\hat H}_e,\hat V]$ being the commutator between $\mathcal{\hat H}_e$ and $\hat V$. Equation~\eqref{timeevolutiongenericoperator} represents the exact time evolution and coupling constant dependence (embedded in $\hat U$) of the 
operator $\hat A$. We consider a perturbation as the sum of $L$ on-site perturbations (which is here referred to as a \emph{global} quantum quench), 
$L$ being the chain length
\begin{equation}
\hat V=\sum_{j=1}^{L}\hat V_j\label{globalquenchoperator} \,\, ,
\end{equation}
where $\hat V_j$ acts on the $j$th site only.

\begin{figure*}[t]
\centering
\includegraphics[width=8.6cm]{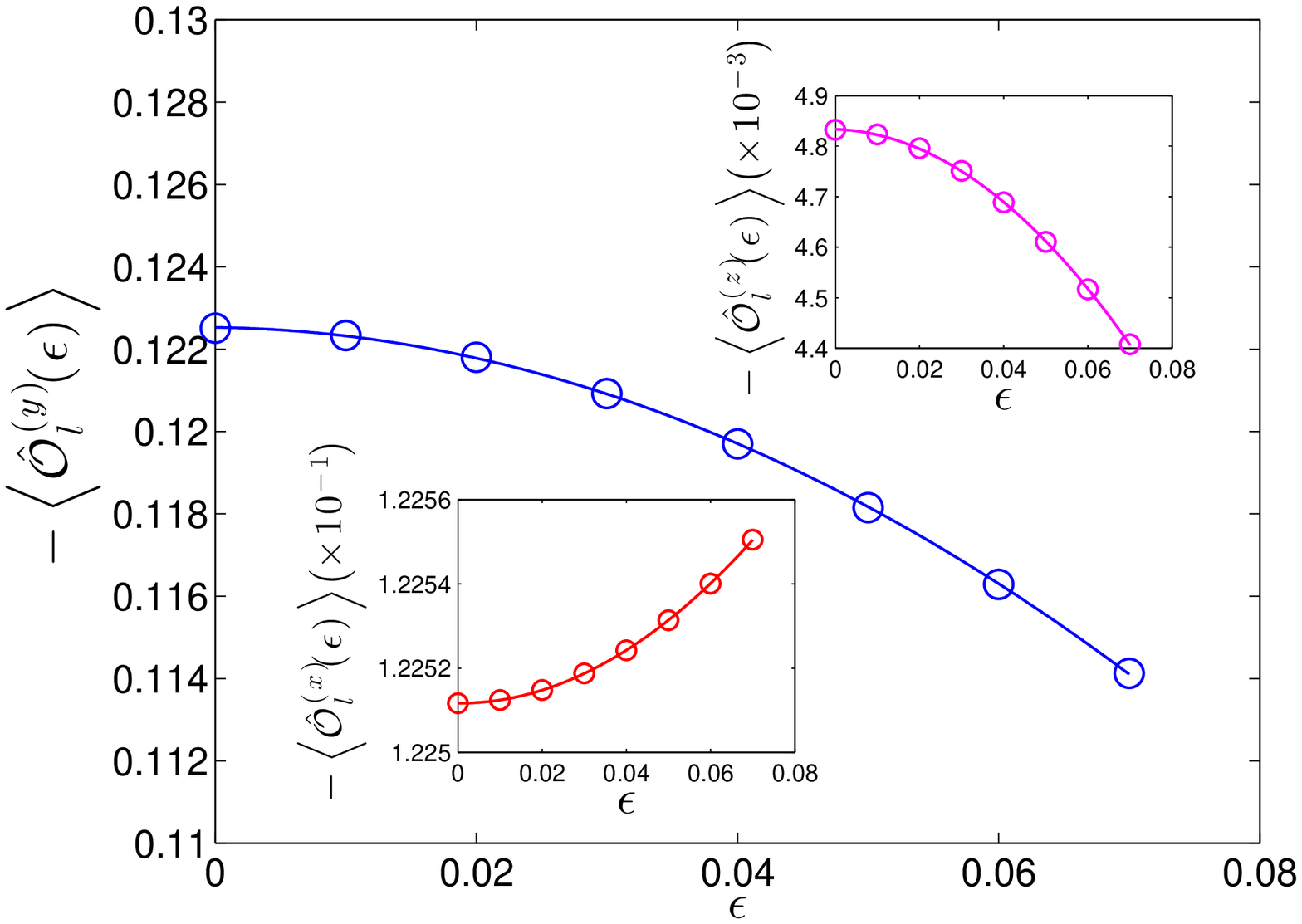}
\includegraphics[width=8.6cm]{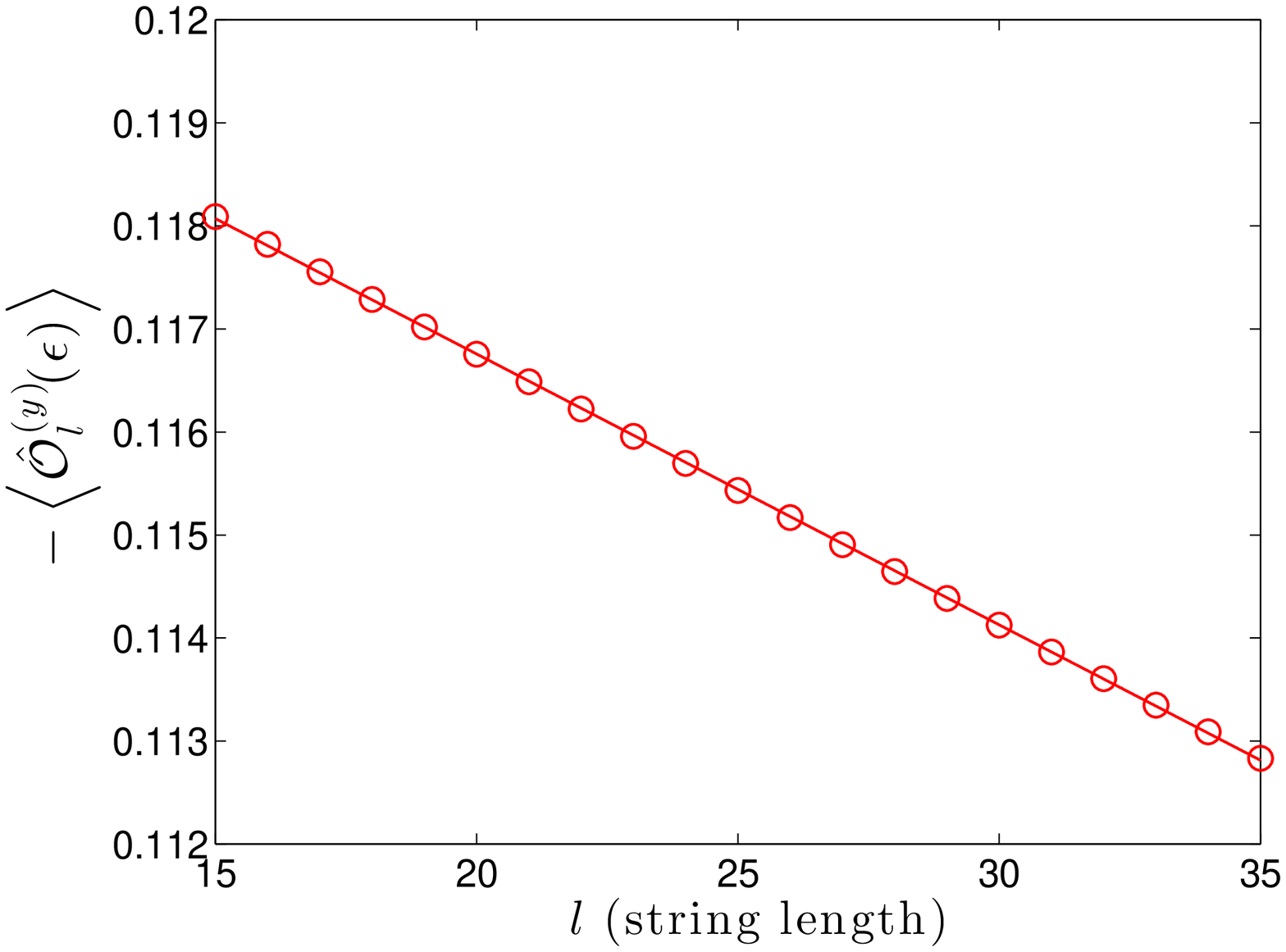}
\caption{(Color online) Numerical simulations for the $\epsilon$ and string-length dependence of the expectation value of the string operator. As in Fig.~\ref{fig:oldsimulationsxxzodd}, the system is quenched with Hamiltonian (\ref{xxzhamiltonian}). We use $L=60$ and $\Delta=0.2$. (Left) Expectation value of the string operator along $y$ as a function of the coupling constant at time $t=1.6\,J^{-1}$ and string length $l=30$ to catch bulk value of the expectation value of the string operator, away from the thermalized region. 
The quadratic behavior predicted by Eq.~\eqref{stringorderintimeandepsilonxxz} is confirmed by the quadratic fit (solid line). A similar behavior is seen for the expectation value of the string operator along $x$ and $z$ (insets). (Right) Expectation value of the string operator along $y$ in the bulk for the perturbation as in the left panel, at time $t=1.6\,J^{-1}$ and coupling constant $\epsilon=0.07$. SO along $y$ (red dots) linearly melts in $l$, as confirmed by the linear fit (red straight line). This is consistent with the fact that an exponential decay is observed away from the perturbative regime.}
\label{fig:quadraticscalingepsilonxxz}
\end{figure*}

\subsection{Quench with the XXZ Hamiltonian in a transverse field}
\label{sec:quenchwiththexxzhamiltonianinatransversefield}

Using Eq.~\eqref{timeevolutiongenericoperator} we are interested in  describing the evolution of the expectation value of string operators 
along the $\alpha$ axis, highlighting the 
dependence on time, coupling constant $\epsilon$, and string length:
\begin{equation}
\left\langle \hat{\mathcal{O}}_l^{(\alpha)} \right\rangle(t,\epsilon)=
\langle\Psi_0|\, \hat U^\dag(t)
\,
\hat{\mathcal{O}}^{(\alpha)}_l(t)
\,
\hat U(t) \, |\Psi_0\rangle \label{stingorderevolution} \,\, .
\end{equation}
We apply perturbation theory to the  model presented in Sec.~\ref{sec:numericalresults}. In what follows we rename by $\epsilon$ the coupling constant $b_x$ in Eq.~\eqref{xxzhamiltonian} and identify $\hat{\mathcal{H}}_e\equiv\hat{\mathcal{H}}_{\rm XXZ}$. 

First, it should be observed that Eqs.~\eqref{stingorderevolution} and~\eqref{timeexpansionoperator} imply that no SO melting arises when quenching the system with $\hat V=\sum_i \hat S_i^{z}$. 
Since $\left[\hat{\mathcal{H}}_{\rm XXZ},\sum_i \hat S_i^{z}\right]=0$, we have that $\hat V(t) = \hat V$ and $\hat U(t) = e^{- i \epsilon\hat V t}$. Additionally, $\sum_i \hat S_i^{z} |\Psi_{0}\rangle=0$ and thus $\hat U(t) \, |\Psi_0\rangle = |\Psi_0\rangle$, from which it follows that $\left\langle \hat{\mathcal{O}}_l^{(\alpha)} \right\rangle(t,\epsilon) = \langle\Psi_0|\,\hat{\mathcal{O}}^{(\alpha)}_l(t)\, |\Psi_0\rangle$. The perturbation, in this case, does not have any effect. This is in agreement with what is discussed by means of generic arguments in Ref.~\cite{rossinimazzafazioprb} and recalled in Sec~\ref{sec:numericalresults}.

In order to gain a better insight for the case in which the condition~\eqref{eq:check} is violated, we write explicitly the time evolution propagator~\eqref{timeevolutionpropagator} up to second order in $\epsilon$:
\begin{eqnarray}
\hat U(t)&\simeq&\hat{\mathbb{I}}-i\epsilon\int_{0}^{t}d\tau\,\hat V(\tau)-\epsilon^2\int_{0}^{t}d\tau_1\,\hat V(\tau_1)\int_{0}^{\tau_1}d\tau_2\,\hat V(\tau_2)\nonumber\\
&\simeq&\hat{\mathbb{I}}-i\epsilon\,\hat \Sigma(t)-\epsilon^2\hat D(t) \,\, ,
\end{eqnarray}
where the single and double time integrals have been indicated with $\hat \Sigma(t)$ and $\hat D(t)$, and are respectively given by
\begin{subequations}
\begin{align}
\hat \Sigma(t&)=\sum_{n=0}^{\infty}\frac{{(i)}^{n}\,{{\rm ad}({\mathcal{\hat H}_e})}^n(\hat V)}{(n+1)!}\,t^{n+1} \,\, ,\\
\hat D(t)&=\sum_{m,n=0}^{\infty}\frac{{(i)}^{m+n}\,{{\rm ad}({\mathcal{\hat H}_e})}^m(\hat V)\,\,{{\rm ad}({\mathcal{\hat H}_e})}^n(\hat V)}{(n+1)!\,m!\,(n+m+2)}\,t^{m+n+2} .
\end{align}
\label{singleanddoubleintegral}
\end{subequations}
The operator in the right-hand-side of Eq.~\eqref{stingorderevolution} becomes
\begin{eqnarray}
\hat U^\dag(t)\hat{\mathcal{O}}^{(\alpha)}_l(t)\hat U(t)\simeq && {\mathcal{O}}^{(\alpha)}_l(t)-i\epsilon\left[\hat{\mathcal{O}}^{(\alpha)}_l(t),\hat \Sigma(t)\right]\nonumber\\
&& +\epsilon^2\left\{\hat \Sigma(t)\hat{\mathcal{O}}^{(\alpha)}_l(t)\hat \Sigma(t)\right.\nonumber\\
&& -\left.\hat{\mathcal{O}}^{(\alpha)}_l(t)\hat D(t)-\hat D^\dag(t)\hat{\mathcal{O}}^{(\alpha)}_l(t)\right\} ,\nonumber\\
\label{operatorforexpectationvalueexpansionxxz}
\end{eqnarray}
where we introduced $\mathcal{O}^{(\alpha)}_l(t)=\left\langle\Psi_0\right|\hat{\mathcal{O}}^{(\alpha)}_l(t)\left|\Psi_0\right\rangle$.
The key result is that the expectation value of the string operator is predicted to vary as
\begin{eqnarray}
\left\langle\hat{\mathcal{O}}_l^{(\alpha)}\right\rangle(t,\epsilon)\simeq&&{\mathcal{O}}^{(\alpha)}_l(t)+\epsilon^2\left\langle \hat \Sigma(t)\hat{\mathcal{O}}^{(\alpha)}_l(t)\hat \Sigma(t)+\right.\nonumber\\
&&-\left.\hat{\mathcal{O}}^{(\alpha)}_l(t)\hat D(t)-\hat D^\dag(t)\hat{\mathcal{O}}^{(\alpha)}_l(t)\right\rangle  .
\label{stringorderintimeandepsilonxxz}
\end{eqnarray}
The proof of Eq.~\eqref{stringorderintimeandepsilonxxz} is postponed to the end of this section. 

The first term in Eq.~\eqref{stringorderintimeandepsilonxxz} represents the unperturbed time evolution, which is the term responsible for thermalization 
at long time and short length that we observed in Fig.~\ref{fig:oldsimulationsxxzodd} and in Ref.~\cite{rossinimazzafazioprb}. The second term instead represents the leading term in the coupling constant describing the action of the odd perturbation $\hat V$: it is the term which introduces melting. Since this is a perturbative expansion, it is not possible to observe the melting phenomenon as a clear scaling $\left\langle \hat {\mathcal O}^{(\alpha)}_l \right\rangle (t) \sim e^{-l}$. It is, however, possible to distinguish the situations in which $\left\langle \hat {\mathcal O}^{(\alpha)}_l \right\rangle (t)$ depends on $l$ (SO melting) from those in which it does not.

Due to the nested-commutator structure arising from the expansion, a necessary condition for SO to melt is that the odd perturbation must not commute with some element of $\mathcal{G}_{\mathbb{D}_2}$~\cite{note1}, i.e.
\begin{equation}
\left[\hat V_n,e^{i\pi \hat S^{\alpha}_n}\right]\neq0\label{microscopicconditionstringordermelting} \,\, ,
\end{equation}
for some $\alpha$. In this case, the expectation value in Eq.~\eqref{stringorderintimeandepsilonxxz} depends on $l$ and describes the onset of melting along the $\alpha$ axis. The relation between SO melting and the symmetry properties of the Hamiltonian was already clarified by Eq.~\eqref{eq:check}. Here, a compatible necessary condition for SO melting is obtained from a different perspective.

However, Eq.~\eqref{microscopicconditionstringordermelting} provides a novel insight: in order to observe SO melting, the quench has to extend over the whole chain as in Eq.~\eqref{globalquenchoperator}, i.e. the quench has to be global. This is necessary in order to ensure the mentioned $l$ dependence of $\left\langle \hat{\mathcal O}^{(\alpha)}_l\right\rangle(t)$.

The necessary condition in Eq.~\eqref{microscopicconditionstringordermelting}
also clarifies the fact that if $\hat V$ belongs to the $\Gamma_\alpha$ representation of $\mathcal{G}_{\mathbb{D}_2}$ ($\alpha = x,y,z$), then SO will not melt along the $\alpha$ axis; indeed, in this case, the necessary condition~\eqref{microscopicconditionstringordermelting} is not valid. 
This result is consistent with the numerical results reported in Sec.~\ref{sec:numericalresults}, where we observed that a magnetic field along $x$ did not lead to SO melting along the $x$ axis.

In Eq.~\eqref{stringorderintimeandepsilonxxz}, we keep only the second-order term in $\epsilon$ and neglect higher orders since the quadratic 
scaling in the coupling constant has been numerically verified, which tells us that the term of order $\epsilon^2$ is indeed the leading term at small coupling constant.  We show the $\epsilon$ dependence in Fig.~\ref{fig:quadraticscalingepsilonxxz} (left). The numerical simulations reported in Fig.~\ref{fig:quadraticscalingepsilonxxz} (right) show that for small $\epsilon$, the dependence of $\left\langle \hat {\mathcal O}^{(y)}_l \right\rangle (t)$ on $l$ is linear. 
This is consistent with the fact that the term of order $\epsilon^2$ in Eq.~\eqref{stringorderintimeandepsilonxxz} actually catches the lowest-order dependence on $l$ of SO melting. 

At this stage, however, the generic structure of the term of order $\epsilon^2$ in 
Eq.~\eqref{stringorderintimeandepsilonxxz} makes the problem of extracting the explicit dependence on $l$ quite difficult to be tackled.
We can overcome this problem by considering small times, since in this limit we can explicitly use the expansion as in Eqs.~\eqref{singleanddoubleintegral}. However, since $\sum_n\hat S^{\alpha}_n|\Psi_0\rangle=0$ for all $\alpha$, we see from Eqs.~\eqref{singleanddoubleintegral} that we have to expand at least to third order in time, which makes the explicit computation of the expectation value in Eq.~\eqref{stringorderintimeandepsilonxxz} tedious.

The proof of Eq.~\eqref{stringorderintimeandepsilonxxz} follows below. The reader uninterested in the technicalities can go directly to Sec.~\ref{sec:pureakltmodel}.

\begin{figure*}[t]
\centering
\includegraphics[width=8.6cm]{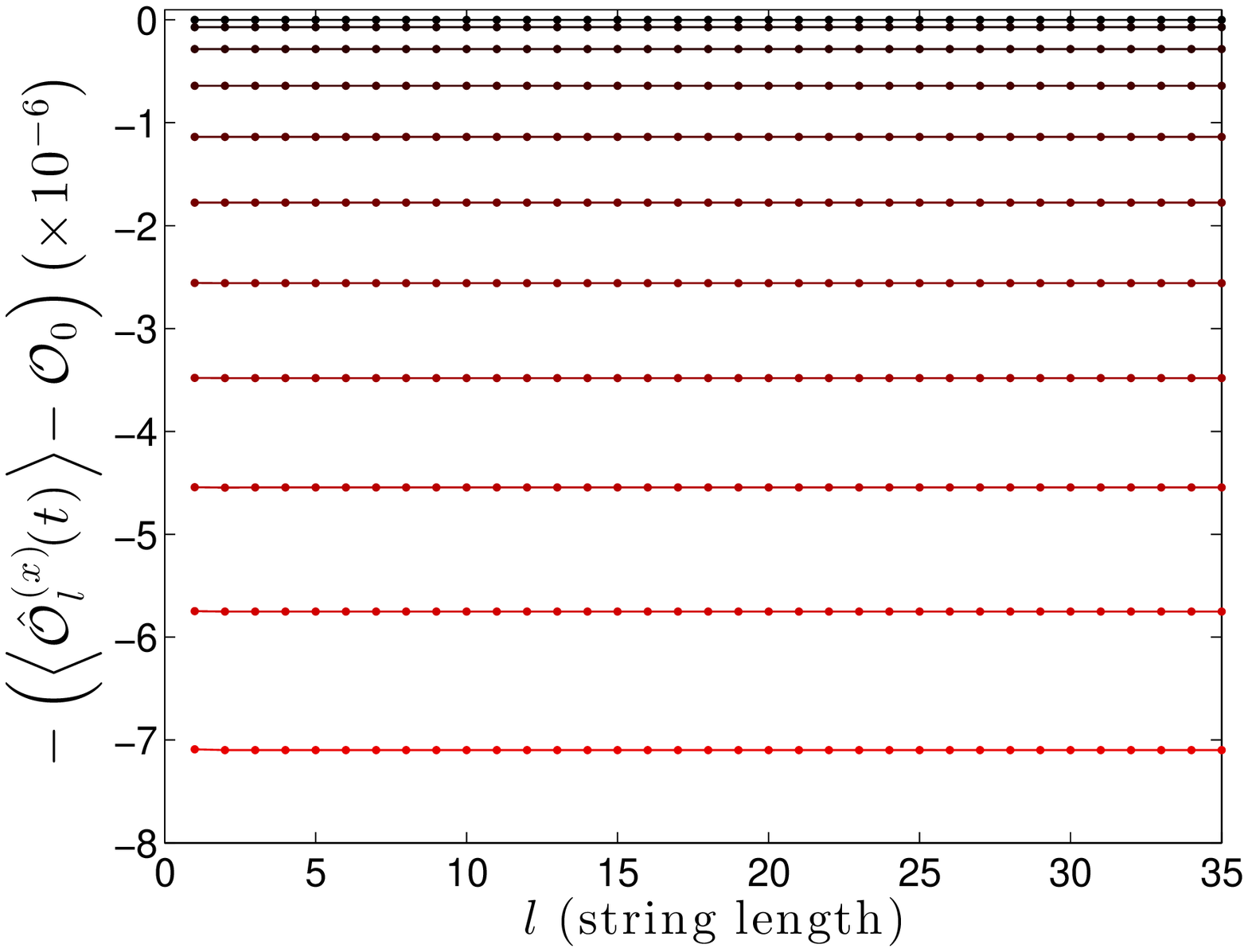}
\includegraphics[width=8.6cm]{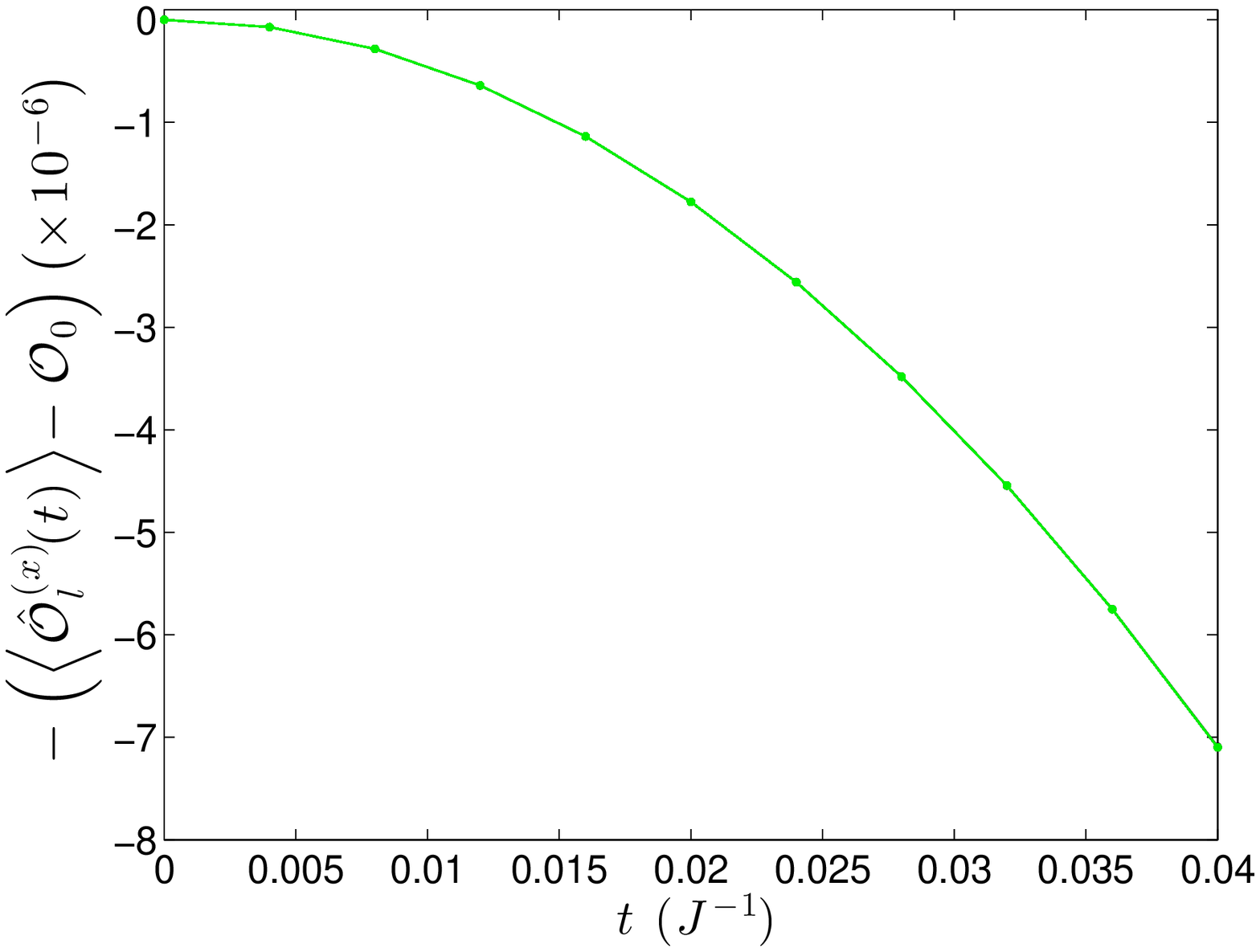}
\includegraphics[width=8.6cm]{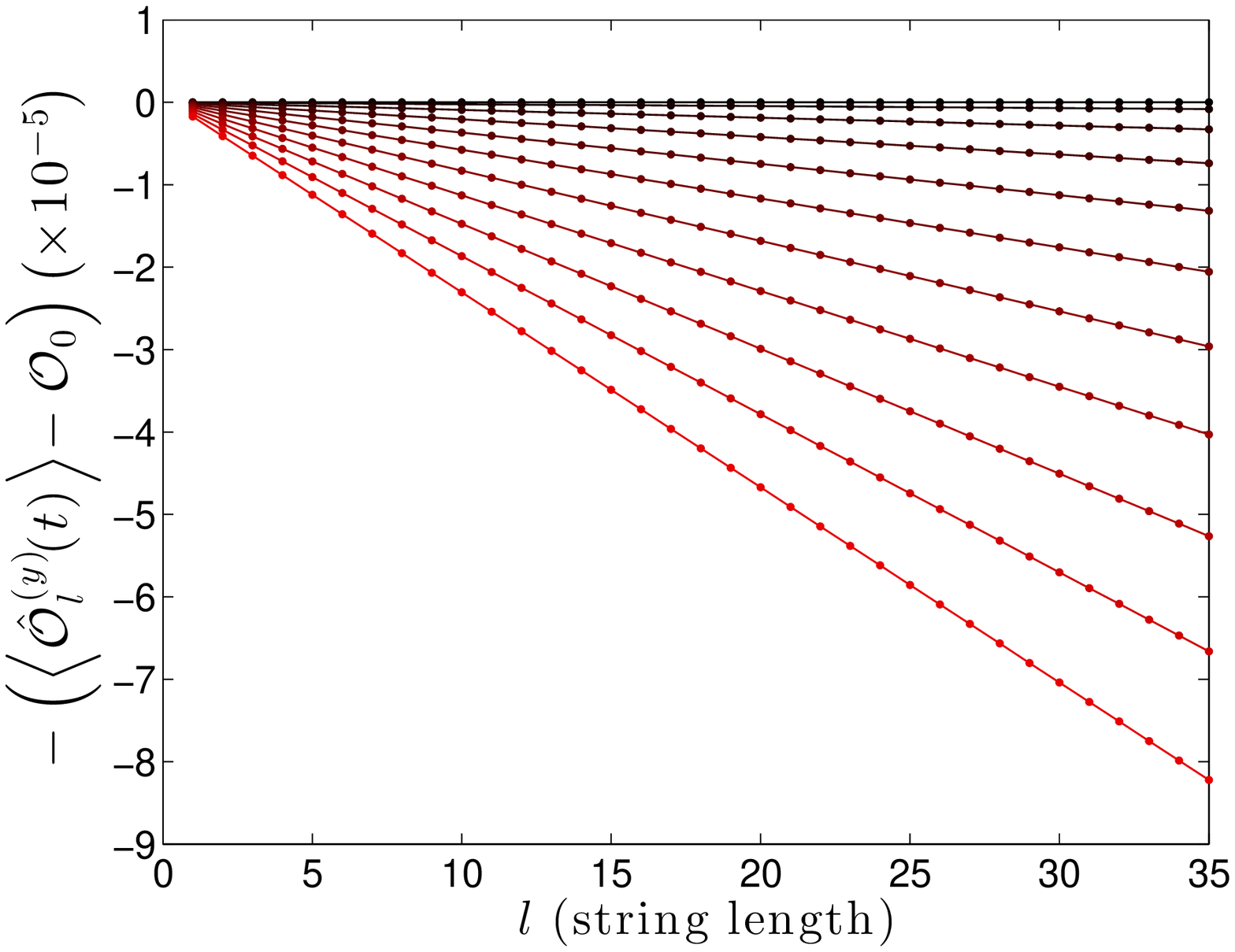}
\includegraphics[width=8.6cm]{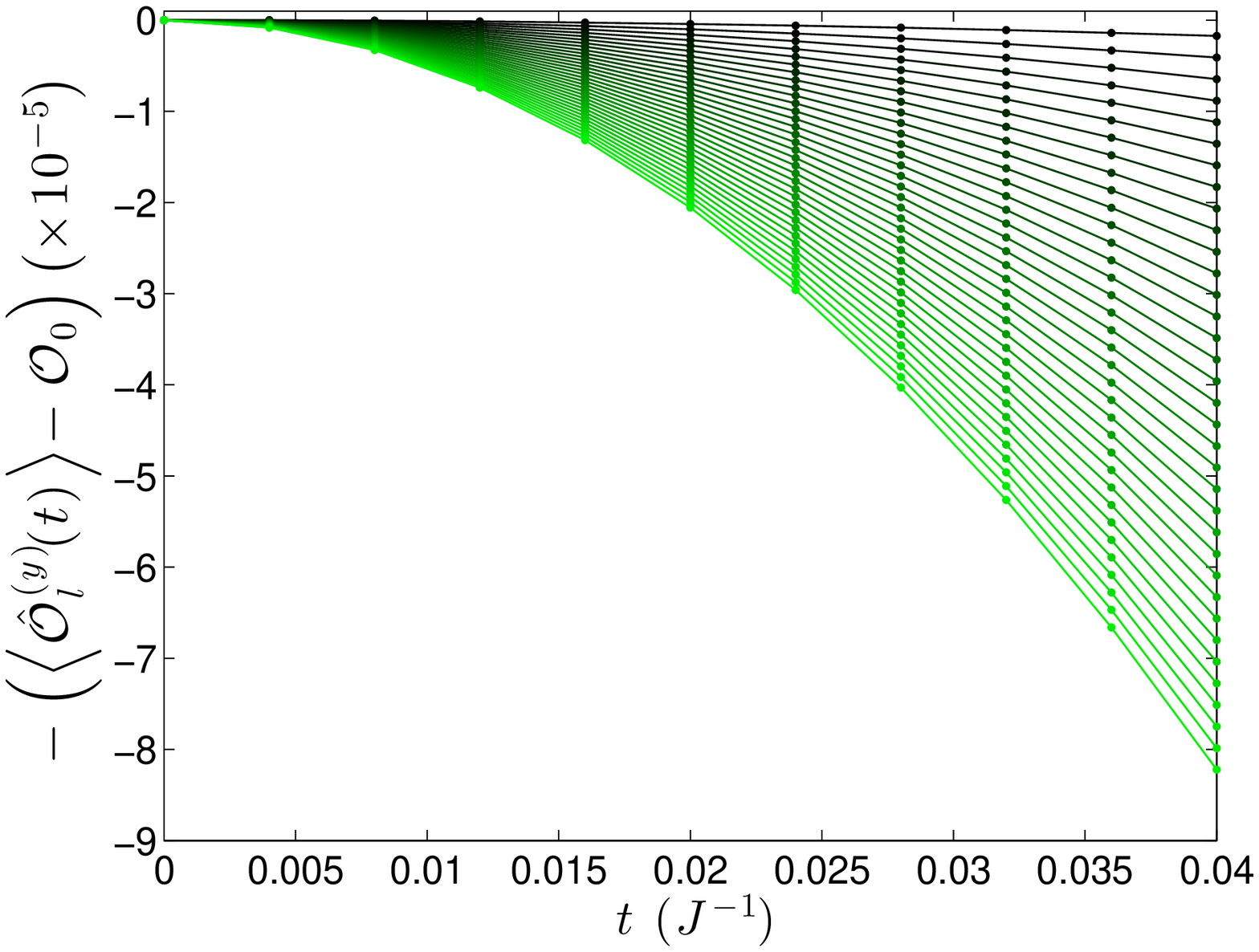}
\caption{(Color online) Quench of the pure AKLT model. Expectation value of the string operator $\left\langle\hat{\mathcal{O}}^{(\alpha)}_l\right\rangle(t)$ for the perturbation $\hat G_x=\sum_i\hat S^y_i\hat S^z_i+\hat S^z_i\hat S^y_i$ at $\epsilon=0.05$ and chain length $L=60$. (Top left) Data for $\alpha=x$ as a function of string length. No dependence on the string length arises as in Eq.~\eqref{expressionstringordernomelting}. Curves are taken at fixed $t$, from $t=0$ (dark red uppermost curve) to $t=0.04\,J^{-1}$ (bright red lowermost curve), with time spacing $\Delta t=0.004\,J^{-1}$. (Top right) Data for $\alpha=x$ as a function of time; curves for different $l$ are superimposed. A quadratic scaling is observed with space-independent curvature as in Eq.~\eqref{expressionstringordernomelting}. (Bottom left) Data for $\alpha=y$ as a function of string length. A linear scaling is observed with time-dependent curvature as in Eq.~\eqref{expressionstringordermelting}. (Bottom right) Data for $\alpha=y$ as a function of time. A quadratic scaling is observed with space-dependent curvature as in Eq.~\eqref{expressionstringordermelting}. Curves are taken at fixed $l$, from $l=1$ (dark green uppermost curve) to $l=35$ (bright green lowermost curve), with length spacing $\Delta l=1$. Color coding in bottom panels is the same as in upper panels.}
\label{fig:datalengthandtimeAKLTGx_xy}
\end{figure*}

\underline{\textit{Proof of Eq.~\eqref{stringorderintimeandepsilonxxz}}}. Since the time evolution is due to $\hat{\mathcal{H}}_{\rm XXZ}$, which is even, any even operator evolved in time by $\hat{\mathcal{H}}_{\rm XXZ}$ remains in the $\Gamma_0$ representation of $\mathcal{G}_{\mathbb{D}_2}$ at any time. The nested commutators, which appear in Eqs.~\eqref{singleanddoubleintegral}, belong to the same $\mathcal{G}_{\mathbb{D}_2}$-representation of $\hat V$. The same is true for $\hat \Sigma(t)$ and $[\hat{\mathcal{O}}^{(\alpha)}_l(t),\hat \Sigma(t)]$ (the string operator is even). On the contrary, $\hat D(t)$ is even by construction. We conclude that, for any odd perturbation $\hat V$, the terms in Eq.~\eqref{operatorforexpectationvalueexpansionxxz} proportional to an odd power of $\epsilon$ do not belong to the $\Gamma_0$ representation.

The previous discussion implies that their expectation value over the AKLT state, which belongs to the $\Gamma_0$ representation of $\mathcal G_{\mathbb D_2}$, vanishes.
The application of an odd operator 
on $|\Psi_0\rangle$ yields a state which is not in the $\Gamma_0$ representation. This state is orthogonal to $\ket{\Psi_{\rm AKLT}}$ by symmetry.
For the same reason, one can generalize this statement to any odd power of $\epsilon$, since in the term of order $\epsilon^{2k+1}$, with $k\in\mathbb{Z}$, the perturbation $\hat V$  appears $2k+1$ times.  $\blacksquare$



\section{Quench of the pure AKLT model}
\label{sec:pureakltmodel}

In order to get a deeper understanding of the melting dynamics, we now present a simplified model for which we are able to describe the behavior of SO in time, string length and coupling constant.

As in Sec.~\ref{sec:stringorderandquantumquenches}, we initialize the system in the AKLT state $|\Psi_{\rm AKLT}\rangle$, but now we 
perform a quench with the Hamiltonian $\hat{\mathcal{H}}=\hat{\mathcal{H}}_{\rm AKLT}+\epsilon\hat V$, where $\hat V$ is an odd perturbation which violates Eq.~\eqref{eq:check}. We are interested in the short time and small coupling constant regime; note that this does not prevent the possibility of observing the physics related to the melting phenomenon, because it is expected to occur at infinitesimal times.

\begin{figure*}[t]
\centering
\includegraphics[width=8.6cm]{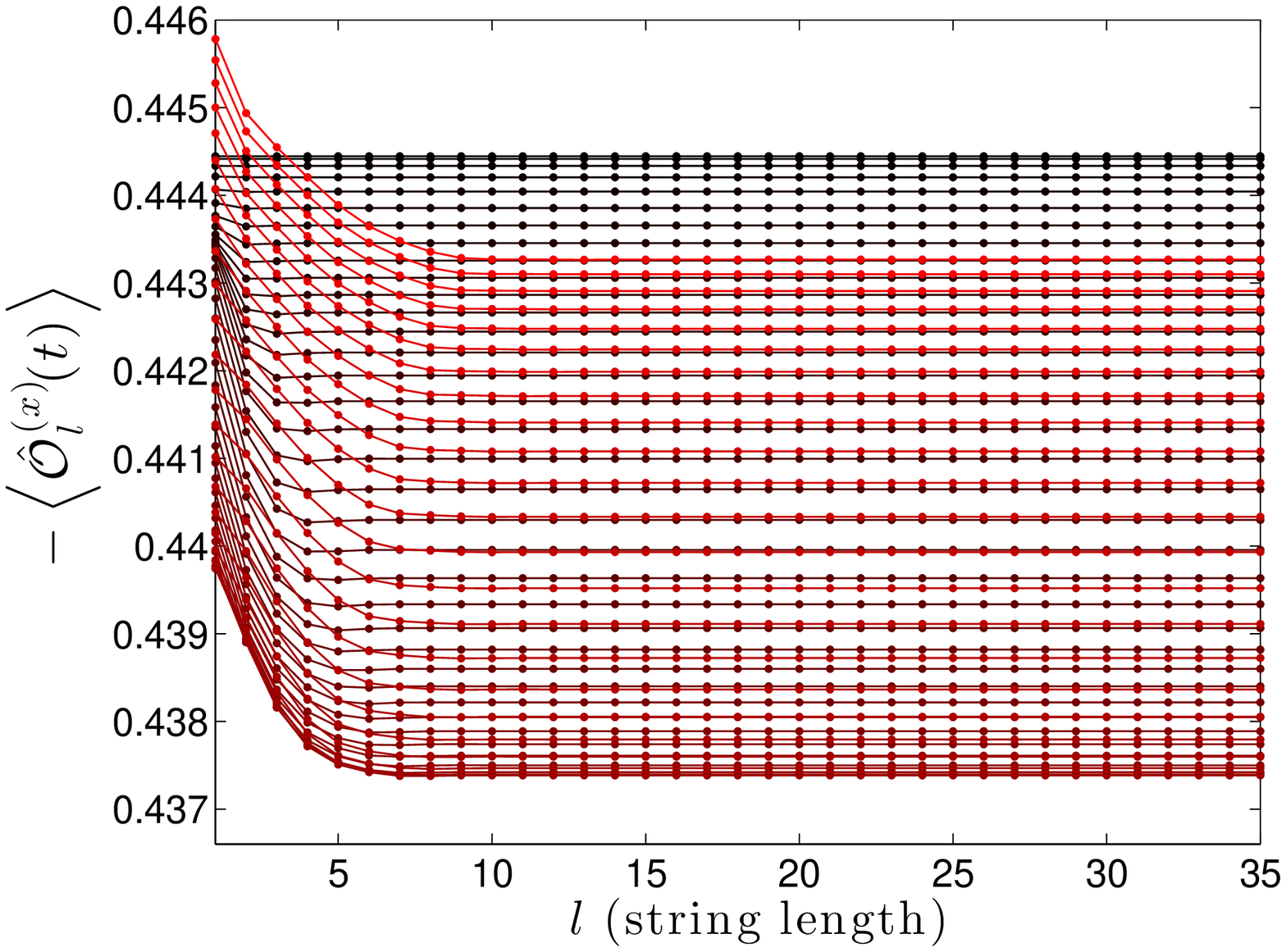}
\includegraphics[width=8.6cm]{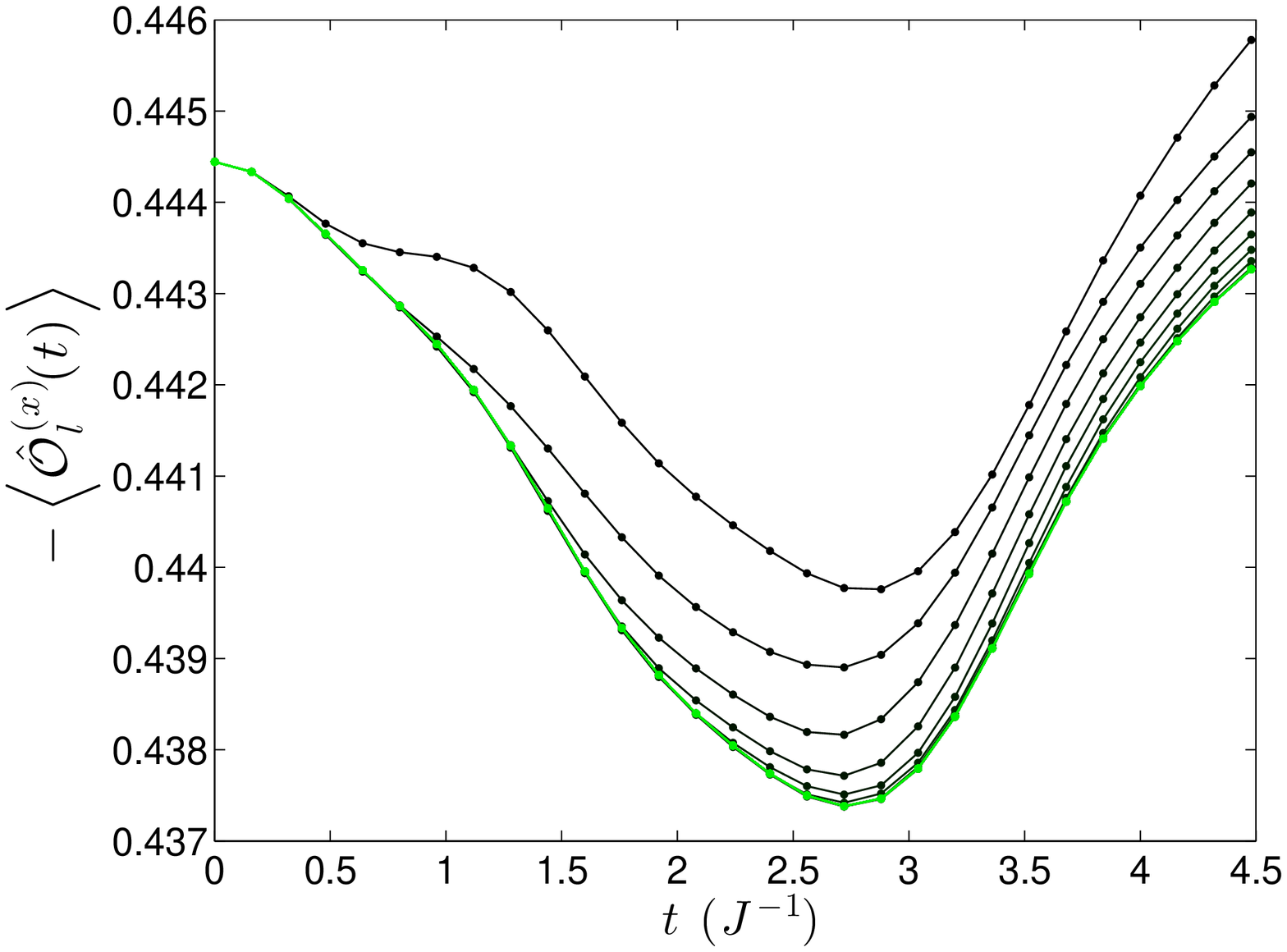}
\includegraphics[width=8.6cm]{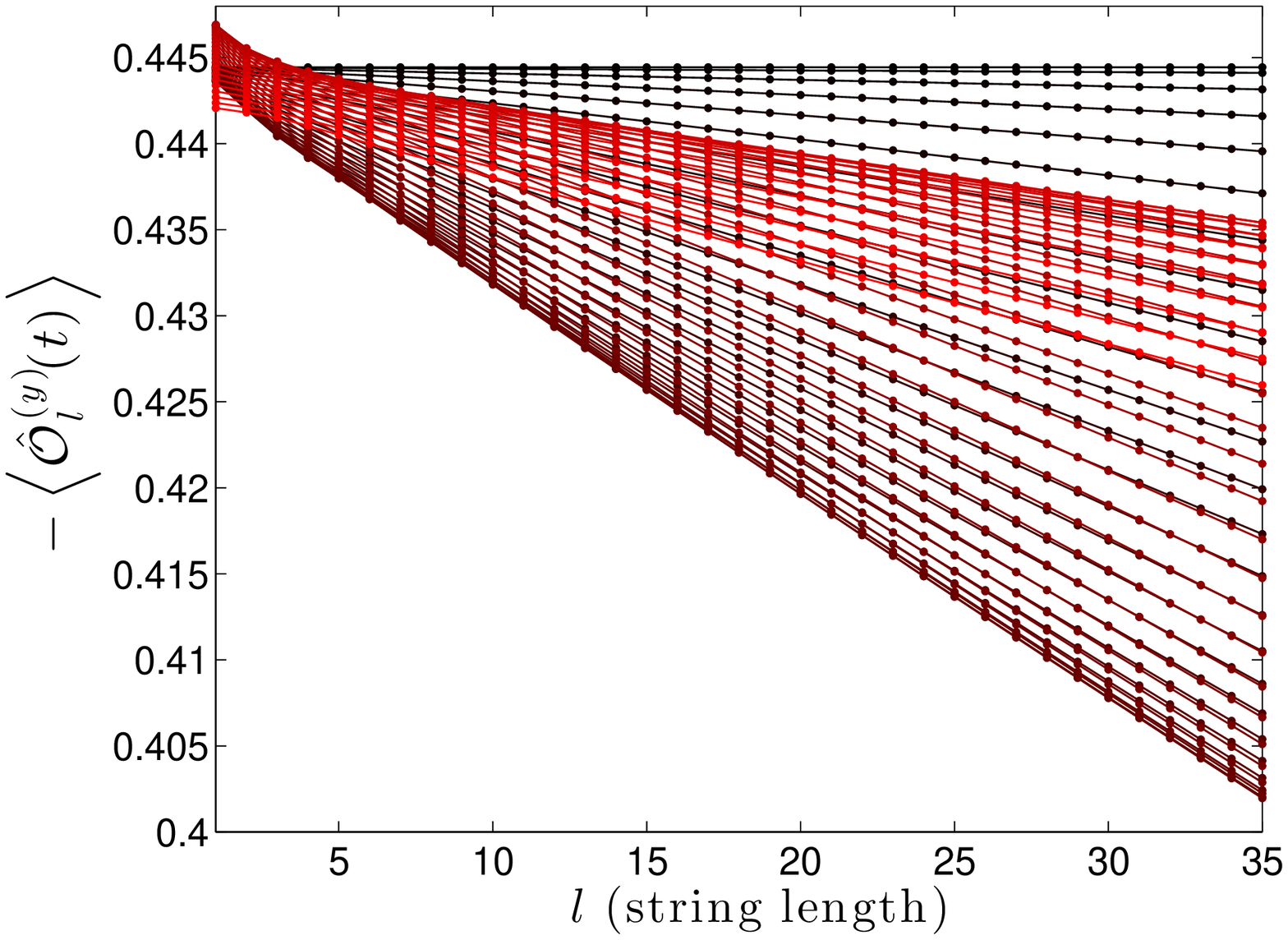}
\includegraphics[width=8.6cm]{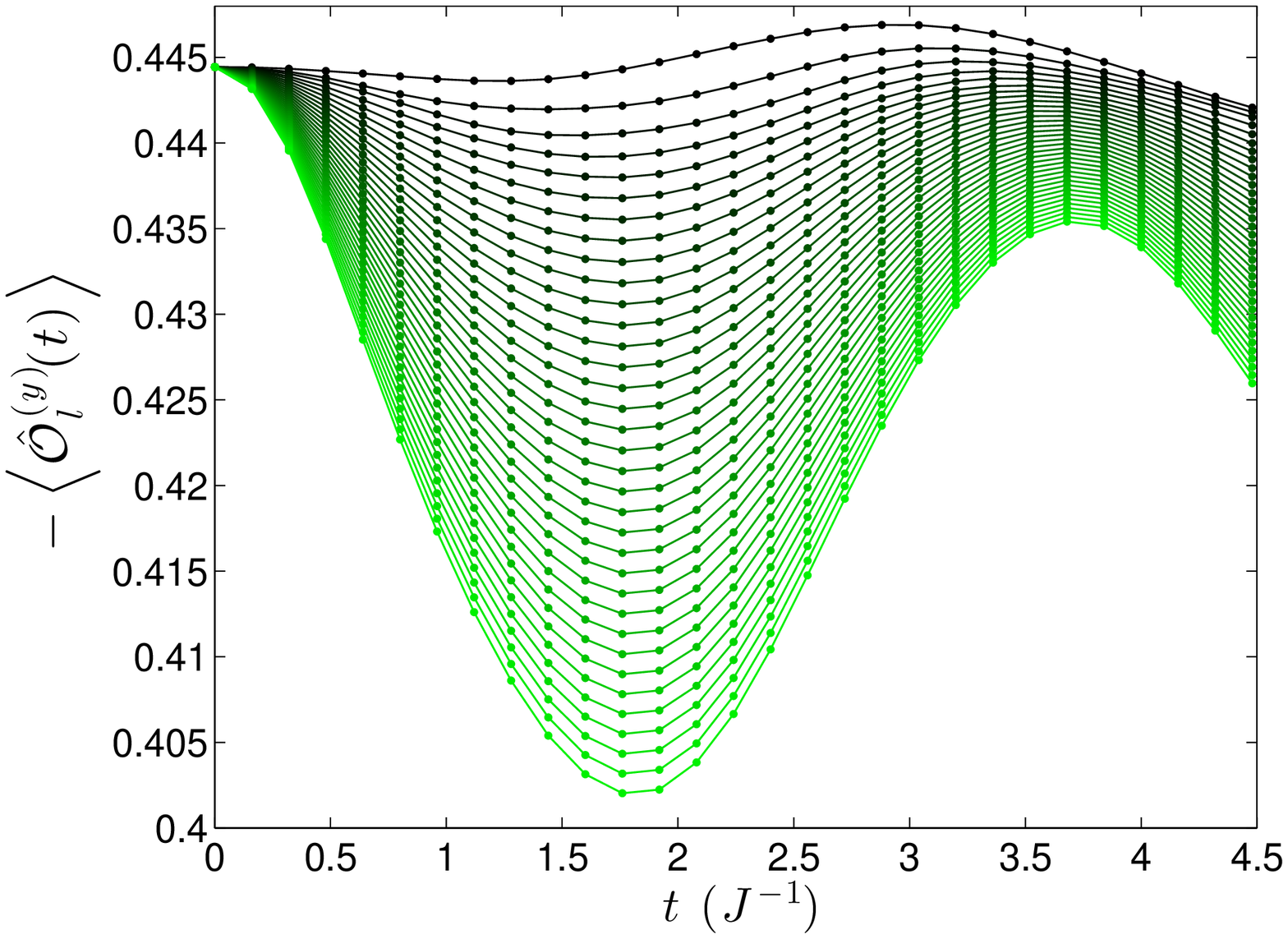}
\caption{(Color online) Same simulation as in Fig.~\ref{fig:datalengthandtimeAKLTGx_xy}, but away from the perturbative regime. We display the long-time behavior of the expectation value of the string operator along the $x$ axis (top panels) and $y$ axis (bottom panels). Curves for string length dependence are taken from $t=0$ (dark red curve, left panels) to $t=4.48\,J^{-1}$ (bright red curve, left panels) by varying $\Delta t=0.08\,J^{-1}$ for each curve. Curves for time dependence (right panels) are taken at fixed string length each, as in Fig.~\ref{fig:datalengthandtimeAKLTGx_xy}. Unfortunately, simulations are not able to reveal a thermalized region, due to the entanglement growth in time.}
\label{fig:datalengthandtimeAKLTGx_xy_long}
\end{figure*}

Since the even part of the Hamiltonian is not changed by the quench, the unperturbed evolution of $\left\langle\hat{\mathcal{O}}_l^{(\alpha)}\right\rangle$ is trivial, and to second order in time, following Eq.~\eqref{stringorderintimeandepsilonxxz}, the expectation value of the string operator varies as
\begin{equation}
\left\langle\hat{\mathcal{O}}_l^{(\alpha)}\right\rangle(t,\epsilon)=\mathcal{O}_0-\epsilon^2\frac{t^2}{2}\left\langle\left[\hat V,\left[\hat V,\hat{\mathcal{O}}^{(\alpha)}_l\right]\right]\right\rangle \,\, ,
\label{stringorderintimeaklt}
\end{equation}
where $\mathcal{O}_0=-4/9$. With this form of perturbation, the dependence on $l$, embedded in $\hat{\mathcal{O}}^{(\alpha)}_l$, can be explicitly extracted. 
By using the expression of the string operator and of the perturbation, one can expand the operator in the expectation value appearing in Eq.~\eqref{stringorderintimeaklt} and separate terms which give a dependence on the string length ($\hat f_l$) from those which do not ($\hat f_0$):
\begin{equation}
\left[\hat V,\left[\hat V,\hat{\mathcal{O}}^{(\alpha)}_l\right]\right]=\hat f_0+\hat f_l \,\, .
\label{operatorexpectationvalueexpansion}
\end{equation}

It is convenient to use perturbations $\hat{G}_\alpha=\sum_i\hat{S}^\gamma_i\hat{S}^\delta_i+\hat{S}^\delta_i\hat{S}^\gamma_i$, where $\alpha\neq\gamma\neq\delta$,  from the orthonormal set as in Table~\ref{tab:tablebasisoperators}. By applying the scheme described before, the following analytical results have been found: if we quench the system with $\hat G_\alpha$, then SO melting occurs along the $\delta\neq\alpha$ axes following the law
\begin{equation}
\left\langle\hat{\mathcal{O}}_l^{(\delta)}\right\rangle(t,\epsilon,l)\simeq\mathcal{O}_0-\frac{1}{2}\epsilon^2t^2\left[X_0+X_1(l-1)\right] \label{expressionstringordermelting}\,\, ,
\end{equation}
where $\mathcal{O}_0=-4/9$, $X_0=-8/9$, and $X_1=-32/27$. Conversely, SO along the $\alpha$ axis does not melt and it evolves according to 
\begin{equation}
\left\langle\hat{\mathcal{O}}_l^{(\alpha)}\right\rangle(t,\epsilon)\simeq\mathcal{O}_0-\frac{1}{2}\epsilon^2t^2X_e \label{expressionstringordernomelting}\,\, ,
\end{equation}
where $X_e=-32/9$ (thus, it is independent of $l$). The proof of Eqs.~\eqref{expressionstringordermelting} and \eqref{expressionstringordernomelting} is postponed to the end of this Section. Away from the perturbative regime an exponential decay of the expectation value of the string operator is observed (see Fig.~\ref{fig:datalengthandtimeAKLTGx_xy_long}). The linear dependence on $l$ found in Eq.~\eqref{expressionstringordermelting}, valid in the perturbative regime, is thus the lowest order in $l$ describing SO melting.

\begin{table}[b]
\centering
\begin{tabular}{l|c|c}
\hline
$\hat M_0^{(1)}=\frac{\hat{\mathbb{I}}}{\sqrt{3}}$ & $\hat M_x^{(1)}=\hat S^{y}\hat S^{z}$ & $\hat M_x^{(2)}=\hat S^{z}\hat S^{y}$\\\hline
$\hat M_0^{(3)}=\frac{3\,\hat S^{z}\hat S^{z}-2\,\hat{\mathbb{I}}}{\sqrt{6}}$ & $\hat M_y^{(1)}=\hat S^{x}\hat S^{z}$ & $\hat M_y^{(2)}=\hat S^{z}\hat S^{x}$\\\hline
$\hat M_0^{(2)}=\frac{\hat S^{x}\hat S^{x}-\hat S^{y}\hat S^{y}}{\sqrt{2}}$ & $\hat M_z^{(1)}=\hat S^{x}\hat S^{y}$ & $\hat M_z^{(2)}=\hat S^{y}\hat S^{x}$\\\hline
\end{tabular}
\caption{Elements of the basis $\mathfrak{B}$. Odd perturbations can be chosen to be $\hat{G}_\alpha=\hat{M}_\alpha^{(1)}+\hat{M}_\alpha^{(2)}$ ($\alpha\neq0$).}
\label{tab:tablebasisoperators}
\end{table}

We show in Fig.~\ref{fig:datalengthandtimeAKLTGx_xy} the results of a simulation with perturbation $\hat G_x$ at $\epsilon=0.05$. In the top panels, we display the expectation value of the string operator along $x$, which is seen to evolve as a function of time and string length according to Eq.~\eqref{expressionstringordernomelting}, whereas the expectation value of the string operator along $z$ behaves as SO along $y$ (not shown). In the bottom panels, we plot the expectation value of the string operator as in the top panels but along the $y$ axis. For small time and string length, the expectation value of the string operator along $y$ is numerically seen to evolve following Eq.~\eqref{expressionstringordermelting}.

In Fig.~\ref{fig:datalengthandtimeAKLTGx_xy_long}, the same simulations as in Fig.~\ref{fig:datalengthandtimeAKLTGx_xy}, up to $t=4.48\,J^{-1}$, are presented. Even if the expectation value of the string operator along the $x$ axis is independent of $l$ in the bulk of the chain (no SO melting), a string-length dependence arises for longer times at short string length (Fig. \ref{fig:datalengthandtimeAKLTGx_xy_long}, top panels) as also observed in Fig.~\ref{fig:oldsimulationsxxzodd}, for the model described in Sec.~\ref{sec:stringorderandquantumquenches}. In the present case, our simulations are, however, still too short to reveal the occurrence of a thermalized region. This is due to the exponential increase of the von Neumann entropy in time, which is in turn related to the bond link dimension of the MPS representation~\cite{uscholwock}.

Notice that these results also apply to the model described in Sec.~\ref{sec:nonlocalorderunderdynamicalperturbations} and Sec.~\ref{sec:analyticalresults} if one uses the $\{\hat G_\alpha\}$ as odd perturbations and expands for small times, since the evolution of the expectation value of the string operator will follow Eq.~\eqref{stringorderintimeaklt} also in this case, but with the only difference that the zero-order term will not be constant and equal to $\mathcal{O}_0$ but will rather vary in time because of the change of the even part of the Hamiltonian, as in Eq.~\eqref{stringorderintimeandepsilonxxz}. This, of course, does not affect the dependence on $l$ and $\epsilon$ in the thermodynamic limit; the predictions given by Eqs.~\eqref{expressionstringordermelting} and~\eqref{expressionstringordernomelting} also hold in this case for SO melting described by the $\epsilon^2$ dependence. This fact has been confirmed by numerical simulations.

\begin{figure}
\centering
\includegraphics[width=8.6cm]{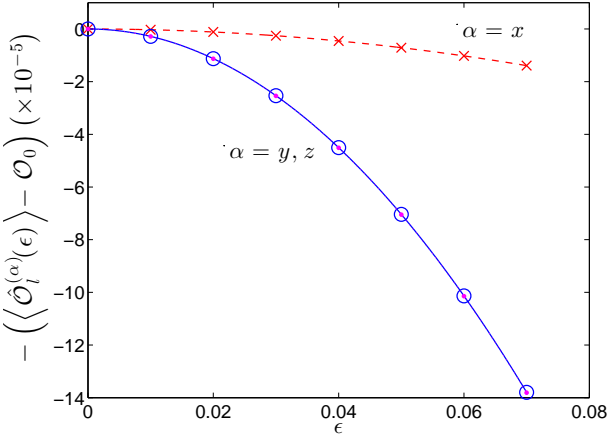}
\caption{(Color online) Quench of the pure AKLT model. Expectation value of the string operator along $x$ axis (red crosses), along the $y$ axis (blue circles), and along the $z$ axis (magenta dots, over the blue circles) as a function of the coupling constant $\epsilon$ for $\hat G_x$ perturbation. In each case data are fitted by a parabola, confirming the quadratic scaling predicted in Eqs. (\ref{expressionstringordermelting}) and (\ref{expressionstringordernomelting}). Data are taken at fixed time $t=0.04\,J^{-1}$ and string length $l=30$ with $L=60$. Numerical values of the curvature extracted by the quadratic fits are: $-0.002839$ ($x$ axis), $-0.02816$ ($y$ axis) and $-0.02819$ ($z$ axis). These values can be compared to those predicted by Eqs.~\eqref{expressionstringordermelting} and \eqref{expressionstringordernomelting} which are: $-0.002844$ ($x$-axis) and $-0.02821$ ($y$ and $z$ axes). Analytical predictions are confirmed.}
\label{fig:dataepsilonAKLTGx}
\end{figure}

A similar comparison can be done for the $\epsilon$ dependence. We show in Fig.~\ref{fig:dataepsilonAKLTGx} numerical data of the expectation value of the string operator along the $x$, $y$, and $z$ axes when we quench with $\hat G_x$ as a function of the coupling constant $\epsilon$. We set the chain length $L=60$ and compute the expectation value of the string operator at fixed time $t=0.04\,J^{-1}$ and, to avoid border effects, we choose $l=30$ since we are interested in bulk values of the string operator. The quadratic behavior in $\epsilon$ predicted in Eqs.~\eqref{expressionstringordermelting} and \eqref{expressionstringordernomelting} is numerically confirmed, as is evident from the picture. We also checked that the curvature, obtained by a quadratic fit of the data, perfectly agrees with the analytical estimates of Eqs.~\eqref{expressionstringordermelting} and \eqref{expressionstringordernomelting} (see the caption to Fig.~\ref{fig:dataepsilonAKLTGx}).
The same results predicted by Eqs. (\ref{expressionstringordermelting}) and (\ref{expressionstringordernomelting}) are found if we quench with $\hat G_y$ or $\hat G_z$, with the difference that SO melting does not occur along the $y$ and $z$ axes, respectively. This is a consequence of the special choice of perturbations that we made. From our data, we see that we are always in the regime such that $\epsilon^2t^2\left\langle\left[\hat V,\left[\hat V,\hat{\mathcal{O}}^{(\alpha)}_l\right]\right]\right\rangle\lesssim\mathcal{O}_0$, which is consistent with the fact that we are developing a perturbative expansion.

\underline{\textit{Proof of Eqs.~\eqref{expressionstringordermelting} and \eqref{expressionstringordernomelting}}}. The terms arising from the expansion of the operator in Eq.~\eqref{operatorexpectationvalueexpansion} which yield a space-independent contribution are
\begin{eqnarray}
\hat f_0=&&\left[\hat V_k,\left[\hat V_k,\hat S^{\alpha}_k\right]\right]\left(\prod_{n=k+1}^{k+l-1}e^{i\pi \hat S^{\alpha}_n}\right)\hat S^{\alpha}_{k+l}+\label{constantterm0}
\nonumber\\
&&+2\left[\hat V_k,\hat S^{\alpha}_k\right]\left(\prod_{n=k+1}^{k+l-1}e^{i\pi \hat S^{\alpha}_n}\right)\left[\hat V_{k+l},\hat S^{\alpha}_{k+l}\right]+\label{constantterm1}
\nonumber\\
&&+\hat S^{\alpha}_k\left(\prod_{n=k+1}^{k+l-1}e^{i\pi \hat S^{\alpha}_n}\right)\left[\hat V_{k+l},\left[\hat V_{k+l},\hat S^{\alpha}_{k+l}\right]\right] \,\, .\label{constantterm2}
\end{eqnarray}

The terms yielding a space dependence are
\begin{eqnarray}
\hat f_l=&&2\left[\hat V_k,\hat S^{\alpha}_k\right]\left(\sum_{i=k+1}^{k+l-1}\left[\hat V_i,e^{i\pi \hat S^{\alpha}_i}\right]\prod_{n\neq i}e^{i\pi \hat S^{(\alpha)}_n}\right)\hat S^{\alpha}_{k+l}+\label{stringterm0}\nonumber\\
&&+2\hat S^{\alpha}_k\left(\sum_{i=k+1}^{k+l-1}\left[\hat V_i,e^{i\pi \hat S^{\alpha}_i}\right]\prod_{n\neq i}e^{i\pi \hat S^{\alpha}_n}\right)\left[\hat V_{k+l},\hat S^{\alpha}_{k+l}\right]+\label{stringterm1}
\nonumber\\
&&+\hat S^{\alpha}_k\left(\sum_{i=k+1}^{k+l-1}\left[\hat V_i,\left[\hat V_i,e^{i\pi \hat S^{\alpha}_i}\right]\right]\prod_{n\neq i}e^{i\pi \hat S^{\alpha}_n}\right)\hat S^{\alpha}_{k+l}+\nonumber\\
&&+\hat S^{\alpha}_k\left(\sum_{i,j=k+1\,:\,i\neq j}^{k+l-1}\left[\hat V_i,e^{i\pi \hat S^{\alpha}_i}\right]\times\right.\nonumber\\
&&\hspace{1.5cm}\times\left.\left[\hat V_j,e^{i\pi \hat S^{\alpha}_j}\right]\,\prod_{n\neq i,j}e^{i\pi \hat S^{\alpha}_n}\right)\hat S^{\alpha}_{k+l}\label{stringterm3} \,\, .
\end{eqnarray}
The space dependence arises only from terms where there is at least one sum: 
space dependence is a consequence of the \emph{nonzero} commutation between the perturbation and the element of $\mathcal{G}_{\mathbb{D}_2}$, as in Eq.~\eqref{microscopicconditionstringordermelting}. 
If $[\hat V_i,e^{i\pi \hat S^{\alpha}_i}]=0$ (which is true if $\hat V_i$ belongs to the $\Gamma_\alpha\neq\Gamma_0$ representation of the $\mathcal{G}_{\mathbb{D}_2}$), the expression in Eq.~\eqref{stringterm3} is zero. 
No space dependence arises and SO along the $\alpha$ axis is preserved.

We can exploit the fact the AKLT state has a simple MPS representation identified by the the set of $2\times2$ real matrices $\left\{A^{(i_k)}\right\}$ to compute all expectation values of the operators $\hat f_0 + \hat f_l$ using the formalism of the \emph{transfer matrices}~\cite{uscholwock} (see Appendix~\ref{appendixcomputationofexpectationvalues}).

To do this, it is convenient to introduce a set of operators forming an orthonormal basis (Hilbert-Schmidt metric) in the space of $3\times3$ complex matrices,
and furthermore belong to a specific representation of $\mathcal{G}_{\mathbb{D}_2}$. A possible orthonormal basis can be $\mathfrak{B}=\left\{\hat M_0^{(1)},\hat M_0^{(2)},\hat M_0^{(3)},\hat M_x^{(1)},\hat M_x^{(2)},\hat M_y^{(1)},\hat M_y^{(2)},\hat M_z^{(1)},\hat M_z^{(2)}\right\}$, where the subscript denotes the symmetry representation. The operators are shown in Table~\ref{tab:tablebasisoperators}. Any operator in the $\Gamma_\alpha=\Gamma_0,\Gamma_x,\Gamma_y,\Gamma_z$ representation of $\mathcal{G}_{\mathbb{D}_2}$ can be expressed as a linear combination of operators in the same representation. Even perturbations can be taken to be $\hat M^{(i)}_0$ for $i=1,2,3$, and odd perturbations can be $\hat G_\alpha=\hat M^{(1)}_\alpha+\hat M^{(2)}_\alpha$ for $\alpha=x,y,z$ to ensure they are Hermitian.
Equations~\eqref{expressionstringordermelting} and \eqref{expressionstringordernomelting} are found using the method described in Appendix~\ref{appendixcomputationofexpectationvalues} and the scheme here explained with perturbations as in Table~\ref{tab:tablebasisoperators}. $\blacksquare$


\section{Conclusions}
\label{sec:conclusions}

In this paper, we have presented an extensive study of  SO melting, namely of the sudden disappearance of SO after a quantum quench. We have highlighted the fact that non-local order can be particularly fragile in a dynamical context, in the presence of a perturbation which breaks the symmetry to which SO is related. This is qualitatively different from what would happen in the presence of a local order parameter in the manner of Landau, where the continuity of the evolution rules out the disappearance of order in an infinitesimal time

We analyzed this phenomenon by considering a quantum quench in a spin-1 chain initialized in the AKLT state. We focused on the coupling constant, time and string-length evolution of the expectation value of the string operator after the quench, and related the presence of SO after the quench to the symmetries of the post-quench Hamiltonian. 

In our simulations, we observed a first exponential decay of the expectation value of the string operator as a function of the string length, which we identify as the thermalization region. This region can be followed either by a plateau (when the system displays SO) or by a second exponential decay.  This second regime is what it is associated to SO melting. A short-time expansion allowed to predict the behavior of SO just after the quench at short times and distances. At longer times, where thermalization occurs, non-perturbative effects come into play and we resorted only on the numerical results.

The features that we studied in our simulations (e.g., light-cone effects) could also be observed when quenching from non-topological states and measuring string correlations (see, e.g., the non-topological Mott insulating phase of the Bose-Hubbard model~\cite{rossinimazzafazioprb} or the fermionic correlations in the Ising chain~\cite{calabreseising}). The melting phenomenon is related to the non-locality of the observable, more than the topology of the state. However, in the AKLT chain, SO is related to the presence of topological order, and thus this out-of-equilibrium analysis can be linked to the survival or disappearance of topological properties in the system.

SO is only one of the features which characterize the properties of one-dimensional symmetry-protected phases of matter~\cite{pollman1, pollman2}. As long as the properties of the ground state are concerned, it is known that different indicators, such as SO, the presence of edge modes or the existence of degeneracies in the entanglement spectrum, may behave differently~\cite{pollman1}.
We leave as an open intriguing perspective the characterization of the behavior of such indicators in an out-of-equilibrium context, as well as the extension of this discussion to higher dimensions. Generally speaking, it would be interesting to assess whether symmetry-protected topological phases can disappear abruptly after a quantum quench.

\begin{acknowledgements}
R.F.~was supported by EU (IP-SIQS and STREP-QUIC), and by Scuola Normale Superiore (progetto interno ``Non-equilibrium dynamics of one-dimensional quantum systems"). 
R.F.~also acknowledges the Oxford Martin School for support.
L.M.~was supported by LabEX ENS-ICFP: ANR-10-LABX-0010/ANR-10-IDEX-0001-02 PSL*. 
L.M.~and D.R.~acknowledge support by the Italian MIUR (FIRB project RBFR12NLNA).
\end{acknowledgements}


\appendix

\section{Proof of Eq.~\eqref{eq:check}}
\label{appendixproofofequation}

We consider the state $\ket{\Psi_0} = \ket{\Psi_{\rm AKLT}}$ and the Hamiltonian $\hat {\mathcal H}_e + \hat {\mathcal H}_o$. We now prove that the time-evolved state $|\Psi(t)\rangle=e^{-i(\hat{\mathcal{H}}_e+\hat{\mathcal{H}}_o)t}|\Psi_0\rangle$  displays SO as defined in Eq.~\eqref{eq:string:order} at infinitesimally short times if and only if Eq.~\eqref{eq:check} is verified. For better readability, we report Eq.~\eqref{eq:check} here below:
\begin{equation}
 \hat {\mathcal H}_o
 \big( \hat {\mathcal H_e} \big)^n
 \ket{\Psi_{\rm AKLT}} = 0\,\, , \quad 
 \forall n \in \mathbb N \,\, .
\end{equation}

\textit{Proof that SO at short time implies Eq.~\eqref{eq:check}.}
We assume that there is a time $t$ such that the state $\ket{\Psi(t)}$ displays SO. Since a necessary condition for a state $|\Phi\rangle$ to possess SO is to be $\mathbb{D}_2$-invariant, i.e. $\hat g|\Phi\rangle=|\Phi\rangle$, for all $\hat g\in\mathcal{G}_{\mathbb{D}_2}$, 
the time-evolved state satisfies $\hat g|\Psi(t)\rangle=|\Psi(t)\rangle$ $\forall\hat g\in\mathcal{G}_{\mathbb{D}_2}$.

We now investigate whether the state $|\Psi(t)\rangle$ can be $\mathbb{D}_2$ symmetric in the presence of an odd Hamiltonian contribution ($\hat{\mathcal{H}}_o\neq0$). 
We consider a small-time expansion: at first order in $t$ one has $|\Psi(t)\rangle\simeq[\hat{\mathbb{I}}-it(\hat{\mathcal{H}}_e+\hat{\mathcal{H}}_o)]|\Psi_0\rangle$. Thus, $|\Psi(t)\rangle$ keeps $\mathbb{D}_2$ symmetry at first order in $t$ if $\hat{\mathcal{H}}_o|\Psi_0\rangle=0$.
One can generalize this by induction to any order in $t$: a full series expansion of $e^{-i(\hat{\mathcal{H}}_e+\hat{\mathcal{H}}_o)t}$ yields that the state $|\Psi(t)\rangle$ keeps $\mathbb{D}_2$ symmetry if and only if
\begin{equation}\hat{\mathcal{H}}_o{(\hat{\mathcal{H}}_e)}^n|\Psi_0\rangle=0 \,\, ,
\end{equation}
for all $n\geq0$, which is indeed Eq.~\eqref{eq:check}. 

Since SO for $|\Psi(t)\rangle$ implies $\mathbb{D}_2$ invariance, it also implies Eq.~\eqref{eq:check}. Note that this proof is valid also for $t$ infinitesimal.

\textit{Proof that Eq.~\eqref{eq:check} implies SO at short time.}
We prove that Eq.~\eqref{eq:check} implies that $\lim_{l \to \infty}\big\langle \hat{\mathcal{O}}^{(\alpha)}_l \big\rangle (t)$ is continuous as a function of $t$. Note that at $t=0$ this expectation value is different from zero. This implies that there is a finite time range where SO is present.

At first order in $t$ one has
\begin{equation}
\left\langle\hat{\mathcal{O}}^{(\alpha)}_l\right\rangle(t)\simeq\left\langle\hat{\mathcal{O}}^{(\alpha)}_l\right\rangle(0)+it\langle\Psi_0|\left[\hat{\mathcal{H}}_e,\hat{\mathcal{O}}^{(\alpha)}_l\right]|\Psi_0\rangle \,\, ,
\label{eq:firsttimeexpansionstringorderforproof}
\end{equation}
where $\hat{\mathcal{H}}_o$ is discarded because of Eq.~\eqref{eq:check} . We assume $\hat{\mathcal{H}}_e$ to be short range, i.e. $\hat{\mathcal{H}}_e=\sum_j\hat{h}_{e,j}$, where $\hat{h}_{e,j}$ acts on $m$ neighboring spins, from $j$ to $j+m-1$. 
The commutator $[\hat{\mathcal{H}}_e,\hat{\mathcal{O}}^{(\alpha)}_l]$ acts at most on $2m$ neighboring spins centered around $k$ and $k+l$.
This follows by explicit inspection: recalling Eq.~\eqref{definitionofstringorder}, we get
\begin{equation}
\left[\hat{\mathcal{H}}_e,\hat{\mathcal{O}}^{(\alpha)}_l\right]=\sum_j\Big[\hat h_{e,j},\hat S_k^{\alpha}
\left(\,
\prod_{n=k+1}^{k+l-1}
e^{i\pi \hat{S}_n^{\alpha}}
\right)
\hat S_{k+l}^{\alpha}\Big] \,\, .
\label{eq:commutator}
\end{equation}
The commutator in Eq.~\eqref{eq:commutator} can be expanded into a \emph{finite} number of terms, around $\hat S^{\alpha}_{k}$ and $\hat S^{\alpha}_{k+l}$, since $\hat h_{e,j}$ is even and commutes with the string, i.e.
\begin{equation}
\Big[\hat h_{e,j},\prod_{n=j}^{j+m-1}e^{i\pi\hat S^{\alpha}_n}\Big]=0 \,\, .
\label{eq:evenhamioltoniancommutaror}
\end{equation}
Because of Eq.~\eqref{eq:evenhamioltoniancommutaror}, the number of terms arising from the expansion of Eq.~\eqref{eq:commutator} is upper bounded by $2m$ also in the limit $l\rightarrow\infty$. 
Moreover, the operator norm of each of these operators is bounded.
It follows that the expectation value of the operator in Eq.~\eqref{eq:commutator} is bounded in the limit $l\rightarrow\infty$; the first-order term in Eq.~\eqref{eq:firsttimeexpansionstringorderforproof} is thus finite in the thermodynamic limit.

A higher-order expansion in time of $\hat{\mathcal{O}}^{(\alpha)}_l(t)$ yields a series of nested commutators between $\hat{\mathcal{H}}_e+\hat{\mathcal{H}}_o$ and $\hat{\mathcal{O}}^{(\alpha)}_l$
\begin{equation}
\left\langle\hat{\mathcal{O}}^{(\alpha)}_l(t)\right\rangle=\sum_{q=0}^{\infty}\frac{{(i)}^q}{q!}\,\left\langle{{\rm ad}\left(\mathcal{\hat H}_e+\mathcal{\hat H}_o\right)}^q\left(\mathcal{\hat O}^{(\alpha)}_l\right)\right\rangle\,t^q \,\, ,
\label{eq:fullnestedcommutatorexpansion}
\end{equation}
where ${{\rm ad}\left(\mathcal{\hat H}_e+\mathcal{\hat H}_o\right)}^q\left(\mathcal{\hat O}^{(\alpha)}_l\right)$ denotes the nested commutator of order $q$ between $\hat{\mathcal{H}}_e+\hat{\mathcal{H}}_o$ and $\hat{\mathcal{O}}^{(\alpha)}_l$, as in Eq.~\eqref{timeexpansionoperator}. Using Eq.~\eqref{eq:check} one can see that, in the expansion~\eqref{eq:fullnestedcommutatorexpansion}, only terms involving $\hat{\mathcal{H}}_e$ are nonzero, and by means of Eq.~\eqref{eq:evenhamioltoniancommutaror} one can prove that the term of order $t^n$ contributes with a finite number of terms in the expansion~\eqref{eq:fullnestedcommutatorexpansion}, for all $n>0$, in the limit $l\rightarrow\infty$. Thus, the term of order $n$ in Eq.~\eqref{eq:fullnestedcommutatorexpansion} is finite in the limit $l\rightarrow\infty$, for all $n>0$, proving that SO is well defined, at least at short times. 

This proves that Eq.~\eqref{eq:check} implies that the state $|\Psi(t)\rangle$ displays SO at short times. Note that this proof is valid also for $t$ infinitesimal. $\blacksquare$

\section{Computation of expectation values}
\label{appendixcomputationofexpectationvalues}
Expectation values are computed thanks to the formalism of transfer matrices. We write the AKLT state as~\cite{klumper, uscholwock} $|\Psi_{\rm AKLT}\rangle=\sum_{\{i\}}\mathrm{Tr}\left[A^{(i_1)}\dots A^{(i_L)}\right]|i_1\dots i_L\rangle$ and define the transfer matrix for a generic on-site operator $\hat \theta$ as
\begin{equation}
\mathbb{E}_{\alpha_n\alpha_{n+1}\beta_n\beta_{n+1}}(\hat \theta)=\sum_{i_n}\sum_{j_n}A^{[i_n]}_{\alpha_n\alpha_{n+1}}A^{[j_n]}_{\beta_n\beta_{n+1}}\theta_{j_ni_n} \,\, ,
\end{equation}
where $\theta_{j_ni_n}=\langle j_n|\hat \theta|i_n\rangle$. The expectation value on the AKLT state of a generic operator $\hat Q$ of the form $\hat Q=\hat B_k\left(\prod_{n=k+1}^{i-1}e^{i\pi \hat S^{\alpha}_n}\right)\hat C_i\left(\prod_{n=i+1}^{k+l-1}e^{i\pi \hat S^{\alpha}_n}\right)\hat D_{k+l}$ is readily computed by means of transfer matrices as
\begin{eqnarray}
\langle\hat Q\rangle=&&\mathrm{Tr}\left[\mathbb{E}(\hat B)\,\mathbb{E}^{i-k-1}\left(e^{i\pi \hat S^{\alpha}}\right)\,\mathbb{E}(\hat C)\times\right.\nonumber\\
&&\times\left.\mathbb{E}^{k+l-i-1}\left(e^{i\pi \hat S^{\alpha}}\right)\,\mathbb{E}(\hat D)\,\mathbb{E}^{L-l+1}(\hat{\mathbb{I}})\right] \,\, .
\end{eqnarray}
In the limit $L\rightarrow\infty$ the expectation value becomes
\begin{eqnarray}
\langle \hat Q\rangle=&&\left\langle1\left|\mathbb{E}(\hat B)\,\mathbb{E}^{i-k-1}\left(e^{i\pi \hat S^{\alpha}}\right)\,\mathbb{E}(\hat C)\times\right.\right.\nonumber\\
&&\times\left.\left.\mathbb{E}^{k+l-i-1}\left(e^{i\pi \hat S^{\alpha}}\right)\,\mathbb{E}(\hat D)\right|1\right\rangle \,\, ,
\end{eqnarray}
where the fact that $\mathbb{E}(\hat{\mathbb{I}})$ is Hermitian and with unitary spectral radius was used, $|1\rangle$ being the eigenvector of $\mathbb{E}(\hat{\mathbb{I}})$ with unitary eigenvalue.

Similarly, the matrices $\mathbb{E}\left(e^{i\pi \hat S^{\alpha}}\right)$ are seen to be Hermitian and with unitary spectral radius for all $\alpha$. They can be then written in the diagonal form $\mathbb{E}\left(e^{i\pi \hat S^{\alpha}}\right)=\sum_{r=1}^{4}\mu_r|\phi_r\rangle\langle\phi_r|$ where $\mu_1=1$ and $\mu_r\equiv\mu=-1/3$ for $r>1$. 
By defining the coefficient $$c_{rs}=\langle1|\mathbb{E}(\hat B)|\phi_r\rangle\langle\phi_r|\mathbb{E}(\hat C)|\phi_s\rangle\langle\phi_s|\mathbb{E}(\hat D)|1\rangle,$$ one can write
\begin{equation}
\langle\hat Q\rangle=\sum_{r,s=1}^{4}c_{rs}\,\mu_r^{i-k-1}\,\mu_s^{k+l-i-1} \label{generalqexpression}\,\, .
\end{equation}
In our case, $\langle \hat Q \rangle$ is independent of $i$ and tends to $c_{11}$ when $l\rightarrow\infty$ since $|\mu|<1$ and $\{c_{rs}\}$ are finite. The same reasoning applies to the other nonlocal operators. 

By using this fact, one can see that the first three terms in Eq.~\eqref{stringterm0} yield a result which scales linearly with $l$. The only term which in principle can give a scaling $\sim l^2$ is the last term in Eq.~(\ref{stringterm0}). For a specific form of perturbation [which means that the $c_{11}$ coefficient in (\ref{generalqexpression}) has to be computable for any form of $\hat B$, $\hat C$ and $\hat D$] a full, quantitative prediction of the evolution of the expectation value of string operator can be given.



\begin{thebibliography}{50}
\bibitem{note1} 
	Recall that, by definition of SO [Eq.~\eqref{definitionofstringorder}] and by means of Eqs.~\eqref{singleanddoubleintegral}, 
	\eqref{timeexpansionoperator} and \eqref{stringorderintimeandepsilonxxz}, the dependence on $l$ is embedded in the string operator.
\end{thebibliography}
\end{document}